\documentclass[12pt,preprint]{aastex}

\usepackage{amssymb}
\usepackage{rotating}
\usepackage{longtable}
\usepackage{lscape}

\newcommand{\HI}{H\,{\sc i}}
\newcommand{\HII}{H\,{\sc ii}}
\newcommand{\kms}{km\,s$^{-1}$}
\newcommand{\vsys}{$v_{\rm sys}$}
\newcommand{\vLG}{$v_{\rm LG}$}
\newcommand{\wtw}{$w_{\rm 20}$}
\newcommand{\wfi}{$w_{\rm 50}$}
\newcommand{\Speak}{S$_{\rm peak}$}
\newcommand{\msun}{M$_{\odot}$}
\newcommand{\fhi}{F$_{\rm HI}$}
\newcommand{\mhi}{M$_{\rm HI}$}
\newcommand{\bgt}{\mbox{$|b| > 10\degr$}}
\newcommand{\blt}{\mbox{$|b| < 10\degr$}}
\newcommand{\AB}{A$_{\rm B}$}
\newcommand{\Ho}{H$_{\rm o}$}

\slugcomment{accepted for publication in the Astronomical Journal}

\shorttitle{The 1000 Brightest HIPASS Galaxies: Newly Cataloged Galaxies}
\shortauthors{Ryan-Weber et al.}

\begin{document}

\title{The 1000 Brightest HIPASS Galaxies: Newly Cataloged Galaxies}

\author{E. Ryan-Weber\altaffilmark{1,2},
B. S. Koribalski\altaffilmark{2},
L. Staveley-Smith\altaffilmark{2},
H. Jerjen\altaffilmark{3},
R. C. Kraan-Korteweg\altaffilmark{4},
S. D. Ryder\altaffilmark{5},
D. G. Barnes\altaffilmark{6},
W. J. G. de~Blok\altaffilmark{2},
V. A. Kilborn\altaffilmark{7,1},
R. Bhathal\altaffilmark{8},
P. J. Boyce\altaffilmark{9},
M. J. Disney\altaffilmark{10},
M. J. Drinkwater\altaffilmark{1},
R. D. Ekers\altaffilmark{2},
K. C. Freeman\altaffilmark{3},
B. K. Gibson\altaffilmark{6},
A. J. Green\altaffilmark{11},
R. F. Haynes\altaffilmark{2},
P. A. Henning\altaffilmark{12},
S. Juraszek\altaffilmark{11},
M. J. Kesteven\altaffilmark{2},
P. M. Knezek\altaffilmark{13},
S. Mader\altaffilmark{2},
M. Marquarding\altaffilmark{2},
M. Meyer\altaffilmark{1},
R. F. Minchin\altaffilmark{10},
J. R. Mould\altaffilmark{13,3},
J. O'Brien\altaffilmark{3},
T. Oosterloo\altaffilmark{14},
R. M. Price\altaffilmark{12,2},
M. E. Putman\altaffilmark{15},
E. M. Sadler\altaffilmark{11},
A. Schr\"oder\altaffilmark{16},
I. M. Stewart\altaffilmark{16,2},
F. Stootman\altaffilmark{8},
M. Waugh\altaffilmark{1},
R. L. Webster\altaffilmark{1},
A. E. Wright\altaffilmark{2},
and M. A. Zwaan\altaffilmark{1}
}

\altaffiltext{1}{School of Physics, University of Melbourne,  
                 VIC~3010, Australia.}
\altaffiltext{2}{Australia Telescope National Facility, CSIRO, 
                 P.O. Box 76, Epping, NSW~1710, Australia.}
\altaffiltext{3}{Research School of Astronomy \& Astrophysics, Mount Stromlo
                 Observatory, Cotter Road, Weston, ACT~2611, Australia.}
\altaffiltext{4}{Departamento de Astronom\'\i{a}, Universidad de Guanajuato, 
                 Apartado Postal 144, Guanajuato, Gto 36000, Mexico.}
\altaffiltext{5}{Anglo-Australian Observatory, 
                 P.O. Box 296, Epping, NSW~1710, Australia.}
\altaffiltext{6}{Centre for Astrophysics and Supercomputing, Swinburne 
                 University of Technology, P.O. Box 218, Hawthorn, VIC 3122,
                 Australia.}
\altaffiltext{7}{Jodrell Bank Observatory, University of Manchester, 
                 Macclesfield, Cheshire, SK11 9DL, U.K.}
\altaffiltext{8}{Department of Physics, University of Western Sydney Macarthur,
                 P.O. Box 555, Campbelltown, NSW~2560, Australia.}
\altaffiltext{9}{Department of Physics, University of Bristol, Tyndall
                 Ave, Bristol BS8 1TL, U.K.}
\altaffiltext{10}{Department of Physics \& Astronomy, University of Wales, 
                 Cardiff, P.O. Box 913, Cardiff CF2 3YB, U.K.}
\altaffiltext{11}{School of Physics, University of Sydney,
                NSW~2006, Australia.}
\altaffiltext{12}{Institute for Astrophysics, University of New Mexico, 
                 800 Yale Blvd, NE, Albuquerque, NM~87131, USA.}
\altaffiltext{13}{WIYN, Inc. 950 North Cherry Avenue, Tucson, AZ, USA.}
\altaffiltext{14}{ASTRON, P.O. Box 2, 7990 AA Dwingeloo, The Netherlands.}
\altaffiltext{15}{CASA, University of Colorado, Boulder, CO 80309-0389, USA.}
\altaffiltext{16}{Department of Physics \& Astronomy,
                 University of Leicester, Leicester LE1 7RH, U.K.}

\begin{abstract}
The \HI\ Parkes\footnote{The Parkes telescope is part of the Australia
Telescope which is funded by the Commonwealth of Australia for
operation as a National Facility managed by CSIRO.} All-Sky Survey
(HIPASS) is a blind 21-cm survey for extragalactic neutral hydrogen,
covering the whole southern sky. The HIPASS Bright Galaxy Catalog
(BGC; Koribalski et al. 2002) is a subset of HIPASS and contains the
1000 \HI-brightest (peak flux density) galaxies. Here we present the
138 HIPASS BGC galaxies, which had no redshift measured prior to the
Parkes multibeam \HI\ surveys.  Of the 138 galaxies, 87 are newly
cataloged. Newly cataloged is defined as no optical (or infrared)
counterpart in the NASA/IPAC Extragalactic Database. Using the
Digitized Sky Survey we identify optical counterparts for almost half
of the newly cataloged galaxies, which are typically of irregular or
magellanic morphological type. Several \HI\ sources appear to be
associated with compact groups or pairs of galaxies rather than an
individual galaxy. The majority (57) of the newly cataloged galaxies
lie within ten degrees of the Galactic Plane and are missing from
optical surveys due to confusion with stars or dust extinction. This
sample also includes newly cataloged galaxies first discovered in the
\HI\ shallow survey of the Zone-of-Avoidance (Henning et
al. 2000). The other 30 newly cataloged galaxies escaped detection due
to their low surface brightness or optical compactness. Only one of
these, HIPASS J0546--68, has no obvious optical counterpart as it is
obscured by the Large Magellanic Cloud. We find that the newly
cataloged galaxies with \bgt\ are generally lower in \HI\ mass and
narrower in velocity width compared with the total HIPASS BGC. In
contrast, newly cataloged galaxies behind the Milky Way are found to
be statistically similar to the entire HIPASS BGC. In addition to
these galaxies, the HIPASS BGC contains four previously unknown \HI\
clouds.
\end{abstract}

\keywords{surveys --- galaxies: distances and redshifts, fundamental
          parameters, kinematics and dynamics --- radio emission lines}

\section{Introduction}
\label{sect:intro}
\setcounter{footnote}{1} The 21-cm line of neutral hydrogen (\HI) is
unique as it can probe regions of the sky where no stars have (yet)
formed (see Schneider 1996). Within individual galaxies, \HI\ is
frequently found well outside the optical radius (e.g. Meurer et
al. 1996; Salpeter \& Hoffman 1996), and many tidal tails or bridges
between galaxies are only detected in \HI\ (see e.g. Koribalski 1996;
Ryder et al. 2001). Until now, the majority of \HI\ observations were
made of objects that had first been identified in the optical (or
lately the infrared), thus imposing \HI\ selection effects on top of
already existing optical selection effects. Important \HI\ structures
like the Leo ring (Schneider 1989) and the Virgo cloud (Giovanelli \&
Haynes 1989) were discovered by accident and indicate the enormous
potential for discovery in an untargeted \HI\ survey.

The sky has been extensively surveyed for galaxies at optical
wavelengths (e.g. Lauberts 1982) but severe limitations remain, mainly
due to the foreground extinction of the Milky Way (which affects
$\sim$25\% of the sky, see e.g. Kraan-Korteweg \& Lahav 2000).  In
many optical catalogs, including the input catalogs for optical
redshift surveys, low-surface brightness (LSB) galaxies are easily
missed and galaxies with diameters less than $\sim$1\arcmin\ are often
misclassified as stars. For example, all objects with brightness less
than 1.15 times the sky and objects classified as stars were excluded
from the input catalog of the Las Campanas Redshift Survey (Shectman
et al. 1996). To supplement the galaxy catalogs, targeted searches for
LSB galaxies (Schneider et al. 1990, 1992; Impey et al. 1996, 2001;
Morshidi-Esslinger et al. 1999; Cabanela \& Dickey 2000) and dwarf
galaxies (Karachentseva \& Karachentsev 1998; Drinkwater et al. 1999;
Drinkwater et al. 2000) as well as deep optical searches for galaxies
behind the southern Milky Way (Woudt \& Kraan-Korteweg 2001) are being
carried out. In the infrared less than $10\%$ of the sky is affected
by foreground extinction, surveys like the Two Micron All Sky Survey
(2MASS, Jarrett et al. 2000) and the Deep Near-Infrared Survey of the
southern sky (DENIS, Epchtein et al.  1997) are now cataloging large
numbers of galaxies.

In contrast \HI\ emission is {\it not} affected by extinction and
enables us to identify many previously hidden galaxies. In addition,
\HI\ can easily be detected in LSB and late-type dwarf galaxies which
are generally gas-rich (Impey \& Bothun 1997).  \HI\ surveys
complement optical galaxy catalogs and substantially improve the
census of galaxies and measurement of the \HI\ content of the local
Universe. \HI\ surveys also clarify voids by placing reliable upper
limits on the mass of objects in these regions.  Furthermore, there
are some components of galaxies that have so far only been discovered
in \HI, e.g. high-velocity clouds (Putman et al. 2002), tidal \HI\
clouds (e.g., HIPASS J0731--69, Ryder et al.  2001) and other nearby
\HI\ clouds (see e.g. Kilborn et al. 2000). Elliptical galaxies, which
are typically \HI-poor, are the main component missing from \HI\
surveys (see e.g. Sanders, 1980; Knapp, Turner \& Cunniffe, 1985)

The current view of the large-scale structure in the Local Universe,
with its filaments and voids, is based almost entirely on optical
observations of high-luminosity galaxies. This view is highly
selective and it will be interesting to see how large-scale surveys
for extragalactic neutral hydrogen affect the current picture. Until
recently, the Arecibo Dual-Beam Survey (ADBS) and the deeper Arecibo
\HI\ Strip Survey (AHISS) were the largest blind 21-cm surveys,
covering areas of 430 and 65 deg$^2$ respectively. Rosenberg \&
Schneider (ADBS, 2000) detected 265 galaxies, of which 81 were
uncataloged, whereas Zwaan et al. (AHISS, 1997; see also Zwaan 2000)
detected 66 galaxies, half of which were uncataloged. With the advent
of a 21-cm multibeam system at the 64-m Parkes telescope (see
Staveley-Smith et al. 1996) as well as new observing and data
reduction software (Barnes et al.  2001), much larger and deeper
surveys are now possible. The \HI\ Parkes All-Sky Survey (HIPASS, see
e.g. Koribalski 2002) is the largest 21-cm survey for neutral hydrogen
to date, covering the whole southern sky. With these surveys,
extragalactic \HI\ astronomy no longer depends entirely on
observations at other wavelengths.

There is a large potential for detecting invisible \HI\ clouds and
uncataloged galaxies with unusual properties in HIPASS. We expect to
find many uncataloged galaxies, either hidden behind the Milky Way
with \HI\ properties similar to the overall galaxy population or
missed due to optical/infrared selection criteria. The former are
important for the completion of our picture of the local large-scale
structure as they bridge previously known structures that are
optically intercepted by the Galactic Plane (e.g. Henning et al. 2000;
Sharpe et al.  2001). The latter are equally important to enhance the
completeness of galaxy catalogs across all morphological types.

Subsets of HIPASS within particular regions of the sky have already
been analysed. In each of these regions uncataloged galaxies were
discovered, many of which are also in our sample. Five uncataloged
galaxies have been found in the Centaurus\,A group (Banks et al.
1999). The South Celestial Cap (SCC) region of the sky ($\delta\ <
-62\degr$) has been studied extensively by Kilborn et al. (2002; see
also Kilborn 2001), who found 114 uncataloged galaxies (out of 536
galaxies in total). Banks et al. (1999) and Kilborn et al. (2002)
searched the HIPASS data to full sensitivity, so only some of their
galaxies will appear in the HIPASS Bright Galaxy Catalog (Koribalski
et al. 2002). On-going analysis of the full-sensitivity HIPASS data
over the total survey area is expected to reveal many more uncataloged
galaxies.  Henning et al. (2000) searched the \HI\ Zone-of-Avoidance
shallow survey (HIZSS; $212\degr\ \leq\ l \leq 36\degr, |b| \leq\
5\degr$) and found 110 \HI\ sources, 67 of which had no published
optical counterpart (see also Staveley-Smith et al. 1998). An \HI\
survey of the Great Attractor region ($l$ = 300\degr\ to 332\degr,
$|b| < 5$\degr) by Staveley-Smith et al. (2000) has so far revealed
305 galaxies, most of which were previously unknown (see also Juraszek
et al. 2000). Complete analysis of deep Zone-of-Avoidance \HI\ data is
under way (Staveley-Smith et al., in prep.).

The \HI\ properties of all 1000 galaxies in the HIPASS Bright Galaxy
Catalog are presented by Koribalski et al. (2002). The optical
properties of all previously catalogued galaxies in the HIPASS BGC are
analysed by Jerjen et al. (2002).  And the \HI\ mass function for the
HIPASS BGC will be discussed by Zwaan et al. (2002). Here we present
the \HI\ properties of 138 galaxies from the HIPASS BGC without
velocity measurements prior to the Parkes multibeam surveys; 87 of
these galaxies are newly cataloged -- that is they do not have a
cataloged optical (or infrared) counterpart listed in the NASA/IPAC
Extragalactic Database (NED). The number distribution of the BGC
galaxies presented in this paper are summarised in
Table~\ref{tab:numbers}. In Section~2 we briefly describe the
observations and the HIPASS BGC selection criteria as well as the
method for optical (and infrared) identifications. In Section~3 we
compare the \HI\ properties of newly cataloged galaxies with low and
high absolute Galactic latitudes. Section~4 contains the
conclusions. A short description of all the newly cataloged galaxies
with identified optical counterparts is given in the Appendix.

\section{Observations \& Selection Criteria}
\label{sect:obs}

The \HI\ Parkes All-Sky Survey (HIPASS) was conducted from 1997 to
2000 with the multibeam system on the 64-m Parkes radio telescope; it
covers the whole southern sky in the velocity range from --1200 to
12700 \kms.  For an overview of the survey parameters as well as data
calibration and imaging techniques see Staveley-Smith et al. (1996)
and Barnes et al. (2001). The HIPASS Bright Galaxy Catalog (BGC,
Koribalski et al. 2002) is a subset of HIPASS and contains the 1000
\HI-brightest sources in the southern sky ($\delta<0^\circ$) based on
their peak flux density (\Speak $\ga$ 116 mJy). Although the total
flux density of a galaxy (\fhi), which relates directly to its \HI\
mass, is a more useful physical measurement, the peak flux density
cutoff was applied as the observations were made in the spectral
domain. Consequently this is not a total flux density limited sample.

The HIPASS BGC selection criteria are briefly described below. The
following velocity ranges were searched for \HI\ signals: $-1200 < v <
-350$ \kms\ and $350 < v < 8000$ \kms, i.e.  omitting the range $|v| <
350$ \kms\ where confusion with high-velocity clouds (see Putman et
al. 2002) makes it difficult to find galaxies.\footnote{Throughout the
paper, the quoted velocities are in the optical convention ($v=cz$)
and heliocentric velocity frame.} Known galaxies with $|v| < 350$ \kms\
as well as HIZSS galaxies (Henning et al. 2000) were added to the
sample. The resulting cutoff, after fitting the \HI\ parameters and
selecting the 1000 brightest \HI\ sources, is \Speak\ $\ga$ 116
mJy. This corresponds to a typical minimum detection level of
9$\sigma$.

\begin{figure}  
\vspace{8cm}
\includegraphics{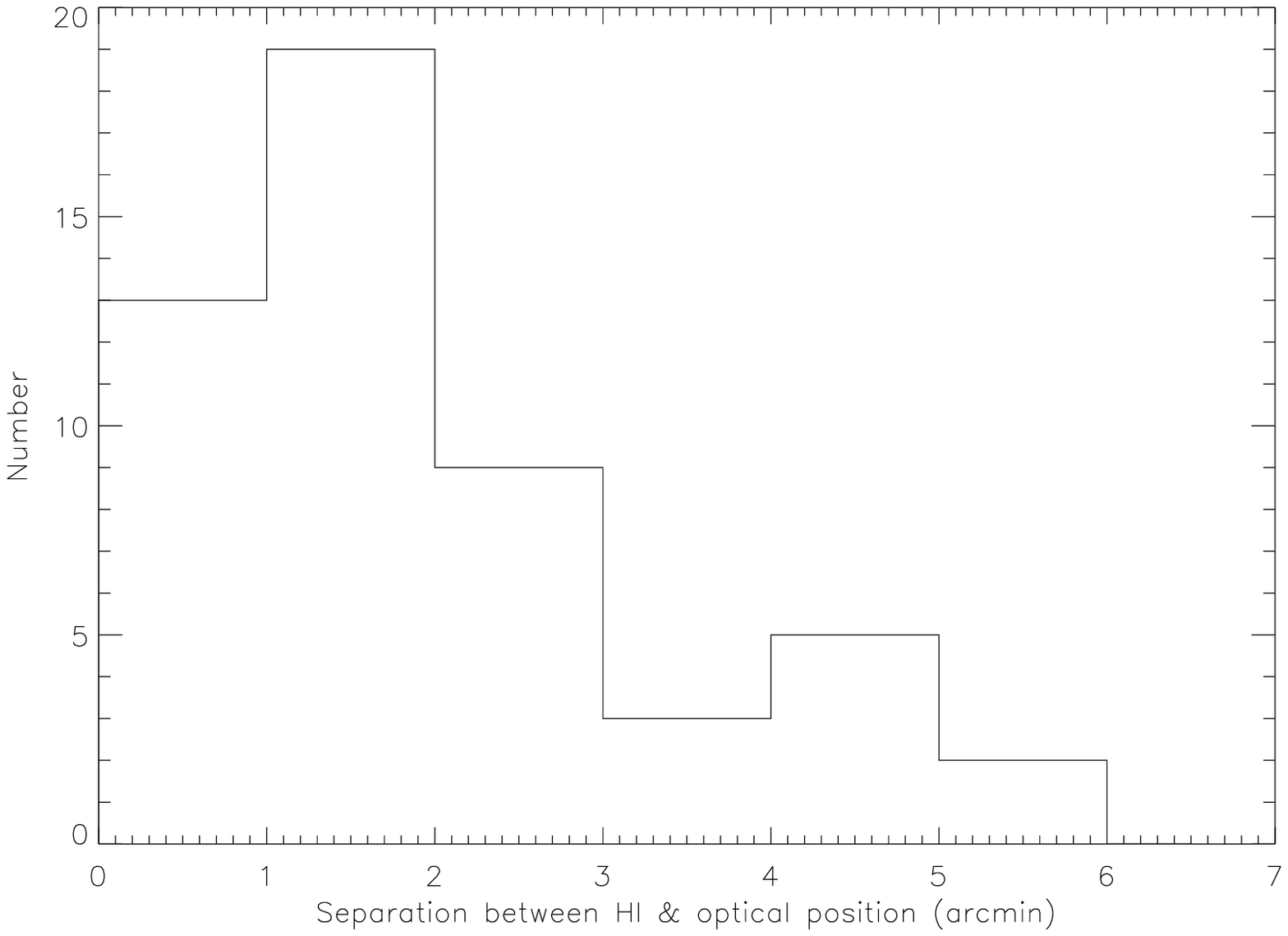}
\caption{\label{fig:sep} Histogram of the angular separations between
         HIPASS and optical position for those newly cataloged
         galaxies in the HIPASS Bright Galaxy Catalog for which
         optical counterparts are identified in the Digitized Sky
         Survey (see Table~\ref{tab:gal}).}
\end{figure}

The galaxy finding process, selection criteria, fitting and analysis
of the \HI\ parameters and identification of cataloged optical
counterparts are described by Koribalski et al. (2002). The FWHM of
the gridded HIPASS beam is 15\farcm5 and the velocity resolution is 18
\kms. The search for optical (and infrared) counterparts was conducted
using NED. We define newly cataloged as any galaxy without an optical
(or infrared) counterpart in NED. For all newly cataloged galaxies,
images from the Digitized Sky Survey (DSS I \& II) were searched for
optical counterparts within an area of 15\arcmin$\times$15\arcmin\
centered on the \HI\ position.  Fig.~\ref{fig:sep} shows the
separations between \HI\ and optical positions. Galaxy positions from
HIPASS are accurate to within a few arcminutes, depending on the \HI\
peak flux density and source extent (Barnes et al. 2001). In addition,
offsets from the optical position can occur intrinsicly due to
multiple optical counterparts, asymmetries or warping of the
\HI. After completion of the identifications with NED late in the year
2000, references to a small number of the newly cataloged galaxies
appeared in the literature, these are noted in
Table~\ref{tab:gal}. Since NED is a dynamic compilation, the
counterparts presented here are valid for NED 2002, March 15th. Some
galaxies may also be present in other catalogs not included in NED
(e.g. The APM Sky Catalogue), these have not been searched.

\section{Results and Discussion}
\label{sect:results}

\begin{figure}  
\vspace{8cm}
\includegraphics{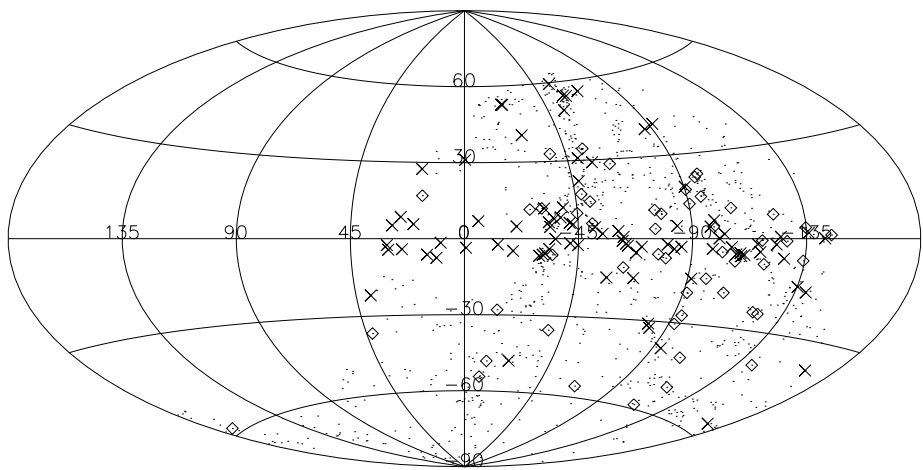}
\caption{\label{fig:lb}Distribution of all galaxies from the HIPASS
	Bright Galaxy Catalog (Koribalski et al. 2002) in Galactic
	coordinates.  The 87 newly cataloged galaxies ($\times$) and
	the 51 known galaxies with no velocity measurements prior to
	the Parkes multibeam \HI\ surveys ($\diamond$) are marked.}
\end{figure}

\begin{figure}  
\vspace{8cm}
\includegraphics{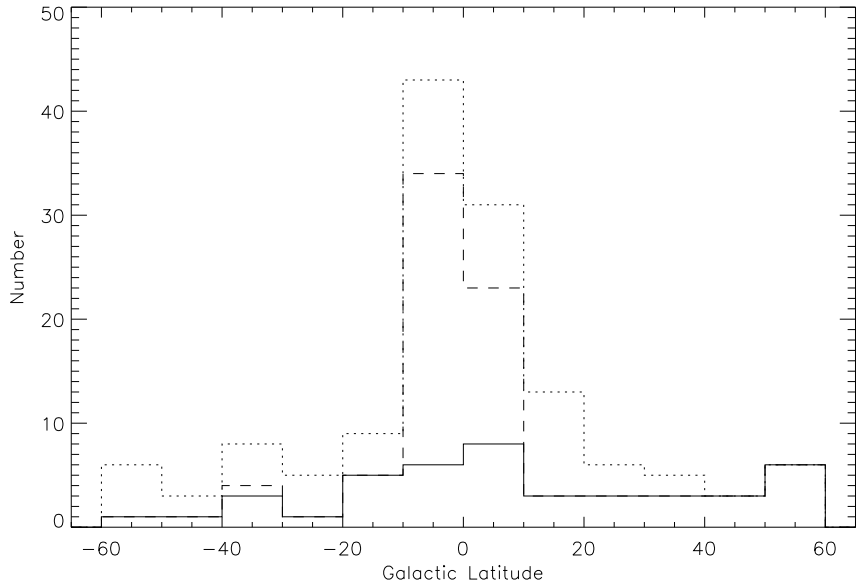}
\caption{\label{fig:histb}Galactic latitude histogram of the 138
  HIPASS BGC galaxies with no previous velocity measurements (dotted
  lines).  The dashed lines mark the subset of 87 newly cataloged
  galaxies, and the solid lines indicate the 43 newly cataloged
  galaxies for which we identify optical counterparts in the Digitized
  Sky Survey.}
\end{figure}

There are 138 galaxies without velocity measurements prior to the
Parkes multibeam \HI\ surveys in the HIPASS BGC. Their distribution on
the sky (Fig.~\ref{fig:lb}) compared to all known HIPASS BGC galaxies,
reveals --- not surprisingly --- most lie near the Galactic
Plane. This is emphasised in the Galactic latitude histogram
(Fig.~\ref{fig:histb}). In the following we concentrate our study on
the 87 newly cataloged galaxies listed in Table~\ref{tab:gal}. For the
analysis we divide the newly cataloged galaxies into two samples:
there are 57 galaxies with \blt\ (see \S~3.1) and 30 galaxies with
\bgt\ (see \S~3.2). We discuss the \HI\ properties of the two samples
and derive optical properties where possible. In \S~3.3 we briefly
discuss the remaining 51 known galaxies without previous velocity
measurements. A short description of the newly cataloged galaxies for
which we have identified optical counterparts is given in the
Appendix. Although \HI\ parameters of the same HIPASS galaxies may
vary slightly between catalogs, depending on the chosen fitting
parameters, the original HIPASS name of each source is maintained for
consistency and cross-identification purposes.

Table~\ref{tab:gal} lists the \HI\ properties of the 87 newly 
cataloged galaxies.
Optical properties are given for those \HI\ sources with one or more
counterparts in the DSS . The columns are as follows:  \\
{\it{ Column}}  1 --- HIPASS Name;  \\
{\it{ Column}}  2 \& 3 --- HIPASS position; \\
{\it{ Column}}  4 \& 5 --- Galactic longitude, $l$, and latitude, $b$, 
                      in degrees;  \\
{\it{ Column}}  6 --- Heliocentric \HI\ systemic velocity, \vsys, in \kms; \\
{\it{ Column}}  7 --- 50\% \HI\ velocity width, \wfi, in \kms;  \\
{\it{ Column}}  8 --- 20\% \HI\ velocity width, \wtw, in \kms;  \\
{\it{ Column}}  9 --- Logarithm of the \HI\ mass, \mhi, in \msun;  \\
{\it{ Column}} 10 \& 11 --- Position of the optical counterpart(s); \\
{\it{ Column}} 12 --- Morphological type within the extended Hubble
classification system (estimated by eye from the DSS); \\
{\it{ Column}} 13 --- References. 

We adopt a uniform percentage error of 10\% on all integrated \HI\
flux densities. This is an empirical estimate based on a comparison of
integrated \HI\ flux densities of 620 galaxies in the LEDA database
(see Koribalski et al. 2002). To calculate the \HI\ mass, recessional
velocities were corrected for the motion of the Sun around the Galaxy
and the motion of the Galaxy in the Local Group. The correction used
is the IAU convention, \vLG\ = \vsys\ + 300 sin($l$) cos($b$). The
\HI\ mass of each galaxy is then calculated using \mhi\ = $2.356
\times 10^5~D^2$~\fhi\ (Giovanelli \& Haynes 1988), where \fhi\ is the
integrated \HI\ flux density in Jy\,beam$^{-1}$\,\kms\ and $D$ =
\vLG\,/\,\Ho\ is the distance in Mpc. We adopt a Hubble constant of
\Ho\ = 75 \kms\,Mpc$^{-1}$.

The HIPASS BGC also contains four \HI\ sources, which are most likely
\HI\ clouds; no obvious optical counterparts have been identified for
these sources, and investigations as to their nature are under
way. Three \HI\ clouds are possibly Magellanic debris (Koribalski et
al., in prep.) and lie within $\sim$10\degr\ of each other, all with
heliocentric velocities $\sim$400 \kms: HIZOA J1616--55 (Staveley-Smith
et al. 1998), HIPASS J1712--64 (Kilborn et al. 2000), and HIPASS
J1718--59 (Koribalski 2001). The fourth cloud, HIPASS J0731--69 (Ryder
et al. 2001), is believed to be a tidal \HI\ cloud associated with the
NGC~2442 galaxy group.

\subsection{Newly Cataloged galaxies with low absolute Galactic
  latitudes (\blt)}
There are 57 newly cataloged galaxies with absolute Galactic latitudes
smaller than ten degrees.  As expected, very few (14) of these have
counterparts in the Digitized Sky Survey; their optical morphologies
range from (Sc) spirals to magellanic irregular (Im) galaxies.  These
galaxies are mainly absent from optical catalogs because of Galactic
foreground extinction. As expected, their \HI\ properties are similar
to the known galaxies in the HIPASS BGC. We expect a small number
($\sim$3\% -- since we find 30 newly cataloged BGC galaxies with \bgt\
out of 1000) of these to suffer from both intrinsic low surface
brightness and Galactic extinction. Indeed, the morphological type of
some newly cataloged galaxies with \blt\ is indicative of this (see
Appendix and Table~\ref{tab:gal}). Spectra of these 57 galaxies are
given in Figure~\ref{fig:hispectra2}.

Thirty seven of the newly cataloged HIPASS BGC galaxies have $|b| <
5$\degr\ (see Fig~\ref{fig:histb}). The \HI\ Zone-of Avoidance shallow
survey (Henning et al. 2000), which is independent from HIPASS,
contains 32 of these galaxies, plus HIZSS\,019 at $b$ = --5\fdg5. The
five additional galaxies are HIPASS J1441--62, J1526--51, J1758--31,
J1812--21, and J1851--09. These were missed in the HIZSS because of
their relatively narrow \HI\ profiles ($w_{\rm 50} < 100$ \kms).  We
identify optical counterparts for only seven of the galaxies with
Galactic latitudes of $|b| < 5$\degr\ (see Table~\ref{tab:gal}). Using
DENIS, Schr\"oder et al.  (1999, 2002) identified at least 14
near-infrared counterparts.

In some cases, optical or infrared counterparts can be seen despite
very high foreground extinction (Schlegel, Finkbeiner, \& Davis
1998). An example is HIPASS J0730--22 (HIZSS\,012; $b$ = --1\fdg9,
\AB\ = 7.8 mag), a spectacular edge-on galaxy with a systemic velocity
of 779 \kms\ (\vLG\ = 528 \kms) and a diameter of $\sim$10\arcmin\ (20
kpc).  We estimated a total dynamical mass within this diameter of
$\sim5 \times 10^{10}$ \msun.

HIPASS J1532--56 (HIZSS\,097, HIZOA J1532--56; $b$ = --0\fdg1) is the
only extended (for definition, see Koribalski et al. 2002) source in
the newly cataloged BGC sample. ATCA \HI\ observations by
Staveley-Smith et al. (1998) show it to be an interacting system.

\subsection{Newly Cataloged galaxies at high absolute Galactic
  latitudes (\bgt)}
There are 30 newly cataloged galaxies with absolute Galactic latitudes
larger than ten degrees. All but one of these galaxies have a
potential optical counterpart. Twenty-five have a single optical
counterpart and four have two or more possible counterparts. The one
galaxy without a possible optical counterpart is HIPASS J0546--68,
which lies behind the Large Magellanic Cloud (LMC). The field is too
obscured to identify an optical counterpart in this case (see Dutra et
al. 2001).

Optical images and \HI\ spectra of 25 newly cataloged galaxies with a
single optical counterpart are shown in Fig.~\ref{fig:images} and
Fig.~\ref{fig:hispectra}, respectively. The four sources with two or
more potential optical counterparts, HIPASS J0605--14, J1225--06,
J1244--08 and J1647--00, are displayed separately in
Figs.~\ref{fig:J0605}, \ref{fig:J1225}, \ref{fig:J1244} \&
\ref{fig:J1647}).

Galaxies have been morphologically classified within the extended
Hubble system set out for giants by Hubble (1926; 1927) and for dwarfs
by Sandage \& Binggeli (1984). The optical morphology of these
galaxies (see Table~\ref{tab:gal}) is dominated by magellanic spiral
and irregular galaxies as well as Blue Compact Dwarf (BCD)
galaxies. In terms of surface brightness we find most galaxies in two
distinct groups: very compact sources of high surface brightness
(e.g., HIPASS J0617--17 and J2200--56) and extended sources of low
surface brightness (e.g., HIPASS J1106--14 and J1255--03). There are
also a few galaxies that have both signatures, a bright core and a low
surface brightness disk (e.g.  HIPASS J1415-04A and J1424--16B). We
conclude that the newly cataloged galaxies with \bgt\ are mostly
absent from optical catalogs because of their small optical diameters
or low surface brightness. The velocity distribution
(Fig.~\ref{fig:vel}) shows that newly cataloged galaxies at high
absolute Galactic latitudes follow the same general trend as all the
newly cataloged galaxies, which is similar to that of the whole HIPASS
BGC (Koribalski et al. 2002).  But their \HI\ mass distribution (see
Fig.~\ref{fig:mass}) is significantly shifted towards lower
values. The median of the mass distribution shifts from
log\,(\mhi/\msun) = 9.4 for newly cataloged galaxies with \blt\ to
log\,(\mhi/\msun) = 8.7 for newly cataloged galaxies with \bgt. A
Kolmogorov-Smirnov test was performed and the distribution of \HI\
masses from the two data sets very found to differ at the 99.6\%
level. The median \HI\ mass of the entire HIPASS BGC is
log\,(\mhi/\msun) = 9.5, similar to that of the newly cataloged
galaxies with \blt.

In Fig.~\ref{fig:curve} we explore the \HI\ profile shapes of the
newly cataloged galaxies by comparing the measured 50\% and 20\%
velocity widths. We find that most of the newly cataloged galaxies at
high absolute Galactic latitudes have narrow \HI\ profiles (mean \wfi\
= 64 \kms). This value stands in sharp contrast to the equivalent
parameter derived for newly cataloged galaxies with \blt\ (135
\kms). A recent survey of local dwarf galaxies (Huchtmeier,
Karachentsev, \& Karachentseva 2001) found a mean \HI\ line width of
\wfi\ = 66 \kms\ (from 98 \HI\ detected dwarf galaxies, velocity
resolution for most galaxies = 6.2 \kms), which is similar to our
value for newly cataloged galaxies with \bgt. Likewise, the standard
deviation of \wfi\ for our sample is 38 \kms\ and Huchtmeier et
al. (2001) find 49 \kms. Correcting for velocity resolution does not
alter these results. Interestingly they find some galaxies with very
narrow profiles (\wfi\ $< 20$ \kms, which would not be found by
HIPASS. Narrow \HI\ velocity profiles are indicative of either face-on
spiral galaxies or low-luminosity galaxies, such as dwarfs. Given that
this catalog is peak flux density selected, future HIPASS catalogs can
be selected by integrated flux density and can contain newly cataloged
galaxies with a different distribution of profiles, including those
from highly inclined spiral galaxies.

There are three galaxies, HIPASS J0403--01, J0605--14 and J1415--04A with
\bgt, for which we measure \wtw\ $\ga$ 200 \kms. The large 20\% velocity 
width of HIPASS J0403--01 (\vsys\ = 910 \kms) is probably due to confusion 
with \HI\ in and around NGC~1507 (= HIPASS J0404--02, \vsys\ = 863 \kms, 
see Koribalski et al. 2002). HIPASS J0605--14 is potentially associated 
with a small group of galaxies. And HIPASS J1415--04A is an edge-on spiral 
galaxy, close to HIPASS J1415--04B.

\subsection{Known galaxies with newly cataloged velocity measurements}
There are 51 cataloged galaxies, in addition to the newly cataloged
galaxies, with no velocity measurement prior to the Parkes multibeam
\HI\ surveys listed in Table~\ref{tab:gal2}. Of these, only 17 lie
within $10^\circ$ of the Galactic Plane.
 
The columns in Table~\ref{tab:gal2} are as follows:  \\
{\it{ Column}}  1 --- HIPASS Name;  \\
{\it{ Column}}  2 \& 3 --- HIPASS position; \\
{\it{ Column}}  4 \& 5 --- Galactic longitude, $l$, and latitude, $b$, 
                      in degrees;  \\
{\it{ Column}}  6 --- heliocentric \HI\ systemic velocity, \vsys, in \kms; \\
{\it{ Column}}  7 --- 50\% \HI\ velocity width, \wfi, in \kms;  \\
{\it{ Column}}  8 --- 20\% \HI\ velocity width, \wtw, in \kms;  \\
{\it{ Column}}  9 --- Logarithm of the \HI\ mass, \mhi, in \msun; \\
{\it{ Column}} 10 --- Galaxy name (from NED, where `c'
         means the galaxy may be confused).\\

Four of the galaxies in Table~\ref{tab:gal2} were previously only classified 
as infrared sources:
\begin{itemize}
\item HIPASS J0747--26 (= HIZSS\,022) is a very faint galaxy associated with
   IRAS\,07451--2610. VLA \HI\ snap-shot observations have been obtained.
\item HIPASS J0809--41 (= HIZSS\,035) is an edge-on galaxy with a diameter of
   $\sim$1\farcm7, identified as the extended source 2MASXi J0809537--414137
   and also known as IRAS\,08081--4132. ATCA \HI\ snap-shot observations
   have confirmed the position.
\item HIPASS J1722--05 is a faint spiral galaxy associated with
   IRAS\,17197--0538.
\item HIPASS J2118--09 is a bright compact galaxy associated with
   2MASXi J2118305--090151 and IRAS\,21158--0914.
\end{itemize}

\subsection{Follow-up Observations}
Follow-up \HI\ observations of the newly cataloged galaxies as well as galaxy
pairs/groups in the HIPASS BGC are under way with the Australia
Telescope Compact Array (ATCA). The aim is to obtain accurate \HI\
positions which will be used to check the candidate optical (and
infrared) counterparts. Some examples are shown in Fig.~\ref{fig:atca}
and Table~\ref{tab:atca}. The table gives the ATCA \HI\ position, the
offset from the HIPASS position, the position angle of the detection
and the total integrated ATCA \HI\ flux density. The offset between
the HIPASS and ATCA position is less than 2\arcmin\ for all galaxies.
In most cases the integrated ATCA \HI\ flux density is $\sim$10-20\%
lower than the HIPASS flux density. This is typical for this type of
observation since an interferometer filters out the more extended,
diffuse \HI\ emission. Such observations are particularly necessary
for confused galaxies where it is not clear which of the optical
counterparts are associated with the \HI\ detection. Numerous \HI\
follow-up observations have also been obtained by Kilborn (2001) and
Kilborn et al. (2002). For example, SCC detection HIPASS J1004--73,
also in our sample, has been observed with the ATCA (Kilborn 2001). It
has a large ($\sim$8 kpc) symmetric disk of \HI\ surrounding the
optical counterpart ($\sim$2 kpc). Follow-up observations with the VLA
have also taken place for some galaxies, e.g. HIPASS J0700--04 (Rivers
2000). Optical spectroscopy is planned to obtain redshifts for some of
the newly cataloged galaxies.

\section{Conclusions} 
\label{sect:con}

A blind \HI\ survey such as HIPASS provides a view of the local
Universe free from optical selection effects. Although the potential
for detecting previously unknown \HI\ structures is high, we do not
find any invisible \HI\ clouds not gravitationally bound to any
stellar system in the HIPASS Bright Galaxy Catalog (BGC). The four
indentified \HI\ clouds are most likely associated with Magellanic
debris or other visible galaxies (The NGC 2442 group in the case of
J0731--69).  This can place important upper limits on the contribution
of \HI\ gas, not associated with galaxies, to the local baryon
density. The HIPASS BGC has improved the census of galaxies in the
local Universe by detecting galaxies behind the Milky
Way. Furthermore, over the whole sky, we have easily detected galaxies
missed in traditional optical surveys due to low surface brightness or
misclassification as stars.

There are 87 newly cataloged galaxies in the HIPASS BGC, an additional
51 galaxies had no redshift measurement prior to the Parkes \HI\
multibeam surveys. The majority (57) of the newly cataloged galaxies
lie behind the Milky Way (\blt) and are missing from optical catalogs
due to confusion or dust extinction. Optical counterparts are found in
the Digitized Sky Survey for only 14 of these galaxies. Statistically,
these 57 galaxies are found to have a similar \HI\ mass distribution
and velocity widths to the entire HIPASS BGC.

All the newly cataloged galaxies with high absolute Galactic latitudes
(30) have a candidate optical counterpart(s) with morphologies ranging
from late-type spiral to irregular, including four with multiple
optical counterparts. The exception is HIPASS J0546--68, which lies
behind the LMC and has no visible optical counterpart. The
characteristic surface brightness of these galaxies is extreme, either
diffuse low surface brightness or compact high surface
brightness. Although these galaxies are \HI-rich, they are not high in
\HI\ mass. The newly cataloged galaxies with \bgt\ on average have a
lower \HI\ mass (median log(\mhi/\msun) = 8.7) and narrower velocity
width (mean \wfi\ = 64 \kms) than \HI\ selected galaxies with
optically cataloged counterparts.

\section*{Acknowledgements}
\begin{itemize}
\item We are grateful to the staff at the ATNF Parkes and Narrabri
      observatories for assistance with HIPASS and follow-up observations.
\item This research has made use of the NASA/IPAC Extragalactic Database (NED)
      which is operated by the Jet Propulsion Laboratory, California Institute
      of Technology, under contract with the National Aeronautics and Space
      Administration.
\item Digitized Sky Survey (DSS) material (UKST/ROE/AAO/STScI) is
      acknowledged.
\item SuperCOMOS Sky Surveys material is also acknowledged.
\end{itemize}

\section*{Appendix}
Here we provide a short description of the newly cataloged galaxies for which
optical counterparts have been identified. In addition to DSS I \& II
we also inspected 2MASS images, where available. Morphologically
classifications have been assigned within the extended Hubble system set out
for giants by Hubble (1926; 1927) and for dwarfs by Sandage \&
Binggeli (1984). The BCDs classified below are only candidates and
will need optical spectroscopy to confirm their morphology.

\subsection*{Newly Cataloged galaxies with low absolute 
Galactic latitudes (\blt)}

The LSB appearance of galaxies in this section is mostly likely due to
foreground extinction. The Galactic foreground extinction in the
photometric B-band, \AB, is estimated from the IRAS DIRBE maps of
Schlegel et al. (1998). Note that the extinction values from the DIRBE
maps are uncalibrated at $|b|<5^\circ$ and may be unreliable.

\noindent {\bf{HIPASS J0718--09}} (HIZSS\,006) must be a low surface
  brightness galaxy, as it is not easily discernible at the relatively
  low extinction level of \AB\ = 1.9 mag. There are 2 extended patches
  of LSB emission visible on the DSS I image which are confirmed on
  the DSS II(R) image; one patch of $\sim$2\farcm0$\times$1\farcm5
  centered on $07^{\rm h}\,18^{\rm m}\,20.8^{\rm s}$,
  --09\degr\,03\arcmin\,20.2\arcsec, and a slightly smaller one of
  $\sim$1\farcm0$\times$1\farcm0 centered on $07^{\rm h}\,18^{\rm
  m}\,14.5^{\rm s}$, --09\degr\,02\arcmin\,59.6\arcsec. Both together
  might define one very extended, face-on LSB source of up to
  4\arcmin. In either case the morphology is hard to determine. Type =
  Sd/Sm.

\noindent {\bf{HIPASS J0730--22}} (HIZSS\,012) is an edge-on, late-type spiral 
  galaxy with a large angular size of $\sim$11\arcmin\ on DSS II(R), not 
  corrected for a Galactic extinction of \AB\ = 7.8 mag. 
  The infrared counterpart is 2MASXi J0730080--220105. For a detailed 
  discussion of the \HI\ and infrared properties of this remarkable galaxy 
  see Hurt et al. (2000). Type = Scd/Sd.

\noindent {\bf{HIPASS J0742--34}} (HIZSS\,019) is a nearly face-on, late-type 
  spiral galaxy with an angular extent of $\sim$1\farcm5$\times$1\farcm0 
  (\AB\ = 6.1 mag). Its infrared counterpart is 2MASXi J0742379--343827.
  Type = Sc/Sd.

\noindent {\bf{HIPASS J0744--35}} is an edge-on spiral galaxy with a distinct
  bulge, very clear on DSS II(R) and 2MASS images (\AB\ = 4.3 mag). Type = Sc.
 
\noindent {\bf{HIPASS J0746--28}} (HIZSS\,021) is a nearby, irregular galaxy 
  with an angular extent of $\sim$40\arcsec$\times$20\arcsec, very clear 
  (\AB\ = 4.3 mag). It is not visible on the 2MASS image. Type = Im.

\noindent {\bf{HIPASS J0833--37}} (HIZSS\,045) is a galaxy with an
  angular extent of $\sim$25\arcsec$\times$20\arcsec\ with a bright
  bulge/nucleus and a small LSB envelope. It seems a bit small for the
  velocity (\vsys\ = 958 \kms) and the extinction (\AB\ = 3.78 mag)
  but could have an obscured LSB halo. Type = Sm? or bulge/nucleus of
  earlier type galaxy.

\noindent {\bf{HIPASS J0904--37}} is an extremely LSB, extended
  (1\farcm75$\times$1\farcm5), face-on spiral dwarf galaxy. It has a bright, 
  small bulge/nucleus and an extended LSB disk (\AB\ = 2.2 mag). Type  = Sc/Sd.

\noindent {\bf{HIPASS J0917--53}} (HIZSS\,053) is an edge-on, irregular galaxy 
  with an angular extent of $\sim$1\farcm0$\times$0\farcm2 (\AB\ = 3.6 mag).
  The surrounding field is very crowded with stars. Type = Sc.

\noindent {\bf{HIPASS J0957--48}} (HIZSS\,060) is a spiral galaxy with
  an angular extent of $\sim$40\arcsec$\times$30\arcsec\ (\AB\ = 2.5
  mag). It consists mainly of a bulge with some LSB halo around it
  (one star very close to the center). The morphology is difficult to
  classify. Type = middle to late-type spiral.

\noindent {\bf{HIPASS J1430--54}} is an extremely LSB face-on spiral
  disk with a very small possible nucleus, visible but even fainter on
  DSS II(R) (\AB\ = 2.9 mag). See also the ATCA image in
  Fig.~\ref{fig:atca}. Type = Sc.

\noindent {\bf{HIPASS J1436--53}} (WKK3285) is a LSB dwarf galaxy
  (\AB\ = 3.4 mag), visible on SRC-J film, very weak on DSS I and
  DSS II(R), roundish, no structure. Its angular extent is
  24\arcsec$\times$17\arcsec, B = 17.7 mag.  See also the ATCA image
  in Fig.~\ref{fig:atca} and Woudt \& Kraan-Korteweg (2001). Type =
  Im.

\noindent {\bf{HIPASS J1451-50}} has a small bright nucleus with a
symmetric outer envelope (\AB\ = 1.4). See also the ATCA
image in Fig.~\ref{fig:atca}. Type = Sm.

\noindent {\bf{HIPASS J1522--49}} (WKK4860) is a galaxy with LSB extended 
  features, a bit clumpy on SRC-J film (\AB\ = 2.6 mag). It is not visible on 
  DSS I and very weak on DSS II(R). 
  Its angular extent is 67\arcsec$\times$20\arcsec, see also Woudt \& 
  Kraan-Korteweg (2001). B = 16.6 mag. Type = Im.

\noindent {\bf{HIPASS J1605--57}} (HIZSS\,101, HIZOA J1605--57,
  WKK5834) is a galaxy with multiple stars superimposed (\AB\ = 2.1
  mag).  See also Juraszek et al. (2000) and Woudt \& Kraan-Korteweg
  (2001). Type = Spiral.

\subsection*{Newly Cataloged galaxies with high absolute Galactic 
latitudes (\bgt)}

\noindent {\bf{HIPASS J0255--10}} is a bright dwarf irregular galaxy
  with one or two bright \HII\ regions, not visible on 2MASS
  images. Type = Im/BCD.

\noindent {\bf{HIPASS J0403--01}} is a LSB galaxy just East of the bright 
  star HD\,25571; it is barely visible on the 2MASS image. Its large 20\% 
  \HI\ velocity width, \wtw\ = 247 \kms, as compared to \wfi\ = 96 \kms\ 
  (see Fig.~\ref{fig:hispectra}) is probably due to confusion with \HI\ in
  and around the galaxy NGC~1507 (= HIPASS J0404--02, \vsys\ = 863 \kms, 
  see Koribalski et al. 2002), located $\sim$20\arcmin\ away.
  Type = Im.

\noindent {\bf{HIPASS J0447--57}} is another LSB galaxy just to the North-West
  of the bright star HD\,30804. It is possibly confused. 
  Type = Im.
 
\noindent {\bf{HIPASS J0532--67}} is an early-type spiral galaxy which lies 
  within the boundaries of the Large Magellanic Cloud (LMC).
  One can recognize a prominent bulge and a low surface brightness disk 
  component. The light distribution is too regular for a BCD (see also the
  2MASS image). This galaxy was also cataloged by Kilborn et al. (2002). 
  The \HI\ position coincides with the infrared sources 2MASXi J0531491--672133
  and IRAS\,05319--6723. Type = Sa or Sb.
 
\noindent {\bf{HIPASS J0605--14}} is associated with a group of galaxies (see 
  Fig.~\ref{fig:J0605}) including two possible LSB optical counterparts. The 
  positions and types of three optical galaxies are given in 
  Table~\ref{tab:gal}.
  The \HI\ spectrum of HIPASS J0605--14 peaks quite sharply between 3000 and
  3100 \kms. Additional low level emission is seen between 3100 and 3200
  \kms. By integrating separately over the two velocity ranges we can 
  associate the bright \HI\ emission with the Im-type galaxy at the center,
  whereas the other two late-type galaxies are probably contained within 
  the lower intensity \HI\ gas envelope to the East. 

\noindent {\bf{HIPASS J0617--17}} is a bright dwarf irregular with one bright
  \HII\ region; it is not visible in the 2MASS data. Type = Im/BCD.

\noindent {\bf{HIPASS J0751--55}} is a spectacular very low surface brightness,
  irregular galaxy close to the stars CD-55\,1980 and CD-55\,1979. 
  It was recently also discovered by Karachentseva \& Karachentsev (2000; 
  [KK2000]~24). Type = Sm/Im.

\noindent {\bf{HIPASS J1004--73}} has a small bulge with smooth
transition into the disk.  Some spiral structure visible in the outer
regions (low surface brightness).  This galaxy was also cataloged by
Kilborn et al. (2002). Type = SBm.

\noindent {\bf{HIPASS J1015--34}} is an \HI\ source close to
  ESO375-G003, but at a lower systemic velocity.  The Nancay \HI\
  spectrum of ESO375-G003 shows a systemic velocity of \vsys\ = 3091
  \kms\ and a velocity width of \wtw\ = 191 \kms\ (Fouqu\'e et
  al. 1990).  Its \HI\ flux density is \fhi\ = 4.5 Jy\,\kms\ with a
  peak flux of $\sim$30 mJy, slightly too faint for a HIPASS
  detection. Interestingly, the Nancay \HI\ spectrum of ESO375-G003
  also includes HIPASS J1015--34. Both sources are part of the IC~2558
  galaxy group (Hopp \& Materne 1985).  ATCA \HI\ observations have
  been obtained. There is a small high surface brightness galaxy
  2\arcmin\ SW of the \HI\ center. Type = BCD.

\noindent {\bf{HIPASS J1106--14}} is a big LSB dwarf Irregular without
  prominent \HII\ regions. It was recently discovered by Karachentsev
  et al. (2000; [KKS2000] 23). Type = Im.

\noindent {\bf{HIPASS J1118--17}} is a compact, high surface brightness galaxy 
  of irregular shape. Type = BCD.

\noindent {\bf{HIPASS J1225--06}} is possibly associated with two
  galaxies (see Fig.~\ref{fig:J1225});
  a high surface brightness dwarf galaxy (LCRS B122316.1--061244) and another
  similar galaxy at $12^{\rm h}\,25^{\rm m}\,39^{\rm s}$, 
  --06\degr\,33\arcmin\,08\arcsec. The \HI\ profile is very narrow, suggesting
  a single galaxy, but there could be additional low level \HI\ emission. 
  Type = Im/BCD. 

\noindent {\bf{HIPASS J1244--08}} could be associated with several galaxies
  (see Fig.~\ref{fig:J1244}), although we note that the \HI\ spectrum shows a
  typical double-horn spectrum indicative of a single gas-rich galaxy. The
  integrated \HI\ distribution shows a point source. The narrow profile either
  indicates a face-on galaxy or a slowly rotating dwarf galaxy as the main 
  component. The full HIPASS spectrum reveals no other \HI\ sources at this
  position. There are at least four galaxies visible in the surroundings of 
  the HIPASS position:
  1) a spectacular, nearly face-on Sm galaxy at 
  $12^{\rm h}\,45^{\rm m}\,13^{\rm s}$, --08\degr\,21\arcmin\,31\arcsec, 
  2) a small, but bright edge-on galaxy, possibly in the background,
  3) an edge-on Sm galaxy at $12^{\rm h}\,45^{\rm m}\,08^{\rm s}$, 
    --08\degr\,23\arcmin\,05\arcsec\ (the infrared counterpart is
    2MASXi J1245078--082305), and
  4) a Im/BCD galaxy at $12^{\rm h}\,45^{\rm m}\,04^{\rm s}$, 
    --08\degr\,23\arcmin\,46\arcsec\ (NPM1G--08.0394). The latter two show
  some signs of interaction.  Numerous small and faint galaxies are 
  visible to the North of this group. \HI\ synthesis imaging is needed to 
  study these galaxies in more detail.

\noindent {\bf{HIPASS J1247--77}} is a nearby irregular, LSB dwarf
 galaxy.  An ATCA \HI\ image has been published by Kilborn et
 al. (2002; their Fig.~15).  HIPASS J1247--77 has the lowest \HI\ mass
 ($\sim 5 \times 10^6$ \msun) among the newly cataloged galaxies in
 both the HIPASS BGC and the SCC sample (Kilborn et al.  2002). Type =
 Im.

\noindent {\bf{HIPASS J1248--08}} is a high surface brightness galaxy just East
  of the bright star HD\,111310. It is also visible in the 2MASS image. The 
  galaxy has a tiny bulge and a strong disk component. Type = late spiral, Sc.

\noindent {\bf{HIPASS J1255--03}} is an LSB dwarf irregular galaxy, not 
  visible in the 2MASS image. Type = Im.

\noindent {\bf{HIPASS J1258--33}} is a late-type galaxy, similar to the LMC.
  Type = SBm.

\noindent {\bf{HIPASS J1300--13B}} is similar to HIPASS J1258--33, except for 
  an LSB extension of the disk to the North.
  Type = SBm(pec). 

\noindent {\bf{HIPASS J1321--31}} is a dwarf irregular galaxy in the
  Centaurus\,A group. It was also discovered by Karachentseva, \&
  Karachentsev (1998; [KK98] 195) and Banks et al. (1999). Type = Im.

\noindent {\bf{HIPASS J1337--39}} is also a dwarf irregular galaxy in the
  Centaurus\,A group (see Banks et al. 1999). Type = Im. 

\noindent {\bf{HIPASS J1415--04A}} is another barred late-type spiral.
  Its infrared counterpart is 2MASXi J1415167--042131 \vsys\ = 2899$\pm$64
  \kms, Colless et al. 2001). The diameter is approximately
  1\farcm1 $\times$ 0\farcm4. Magnitude = 15. Type = SBd.
 The galaxy, HIPASS J1415-04B (see below), is a close neighbour
  (separation = 18\farcm6 or 190 kpc). 

\noindent {\bf{HIPASS J1415--04B}} is a barred Sb or Sc galaxy. It has also
  recently been discovered by Colless et al. (2001; 2dFGRS N145Z235, \vsys\
  = 2880$\pm$89 \kms). A second galaxy, 2dFGRS N145Z228, closer to the \HI\
  position has a much higher velocity of 16912 \kms. The diameter is 
  approximately 0\farcm8 $\times$ 0\farcm7. Magnitude = 14.8. Type = SBb/c. 
 
\noindent {\bf{HIPASS J1424--16B}} is a late-type spiral galaxy. No
  bar or bulge is visible on the DSS II(R) image, but there is some
  evidence for a disk.  Type = Sm/Im.

\noindent {\bf{HIPASS J1434--47}} is a  very LSB dwarf galaxy in a crowded
  field of stars. See also the ATCA image in Fig.~\ref{fig:atca}.
  Type = Im.

\noindent {\bf{HIPASS J1513--44}} is a small galaxy. It appears too bright 
  for an Im galaxy. Type = BCD/Im.

\noindent {\bf{HIPASS J1558--10}} is a dwarf galaxy. Type = Sm/BCD.

\noindent {\bf{HIPASS J1647--00}} is associated with a group of galaxies (see
  Fig.~\ref{fig:J1647} at the center of the \HI\ detection: a peculiar looking
  merged galaxy pair of type Sm at $16^{\rm h}\,47^{\rm m}\,59^{\rm s}$,
  --00\degr\,22\arcmin\,59\arcsec, another Spiral at 
  $16^{\rm h}\,48^{\rm m}\,10^{\rm s}$, --00\degr\,21\arcmin\,48\arcsec\
  and an edge-on Sd at $16^{\rm h}\,47^{\rm m}\,59^{\rm s}$,
  --00\degr\,19\arcmin\,47\arcsec\ (see also Table~\ref{tab:gal}).

\noindent {\bf{HIPASS J2020--04}} is a late-type spiral galaxy. Type = Sm/Im.

\noindent {\bf{HIPASS J2200--56}} is confused. The surrounding field shows a 
  galaxy group or cluster in the background. The \HI\ source is most likely 
  associated with the galaxy APMUKS(BJ) B215715.27--564246.0 (Maddox et al.
  1990) just to the West of the bright double or multiple star HD\,208877.  
  Type = BCD.


\begin{figure}  
\caption{\label{fig:images}DSS images (5\arcmin $\times$ 5\arcmin) of
  the 25 newly cataloged galaxies with high absolute Galactic
  latitudes (\bgt) and a single candiate optical counterpart. Each DSS
  image is centered on the optical position.}
\end{figure}

\begin{figure}  
\caption{\label{fig:hispectra} \HI\ spectra of the newly cataloged
  galaxies with high absolute Galactic latitudes (\bgt) in the HIPASS
  BGC. The Figure location of each spectrum corresponds to the DSS
  image in Fig.~4. In addition we show the \HI\ spectrum of the galaxy
  HIPASS J0546--68 (bottom left), which is obscured by the LMC.}
\end{figure}

\begin{figure}  
\vspace{5cm}
\includegraphics{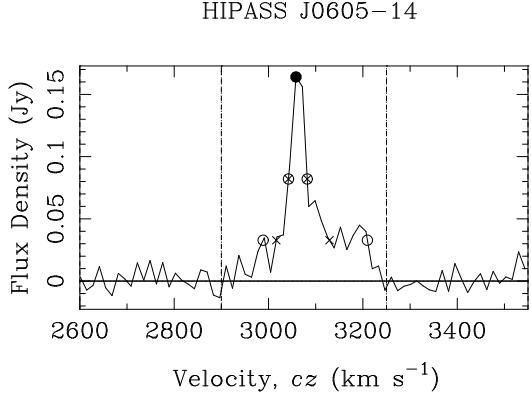}
\caption{\label{fig:J0605} a) \HI\ spectrum of HIPASS J0605--14. b)
DSS image centered on HIPASS J0605--14. There are three potential
optical counterparts. By integrating separately over the two velocity
ranges (grey contours: 3000--3100 \kms; black contours: 3100--3200
\kms) we can associate the bright \HI\ emission with the Im-type
galaxy near the center, whereas the other two galaxies are probably
contained within the lower intensity \HI\ envelope to the East.  The
contour levels are at 60, 70, 80 and 90\% of the maximum \HI\ flux
density (7.8 Jy\,beam$^{-1}$ \kms).}
\end{figure}

\begin{figure}  
\vspace{5cm}
\includegraphics{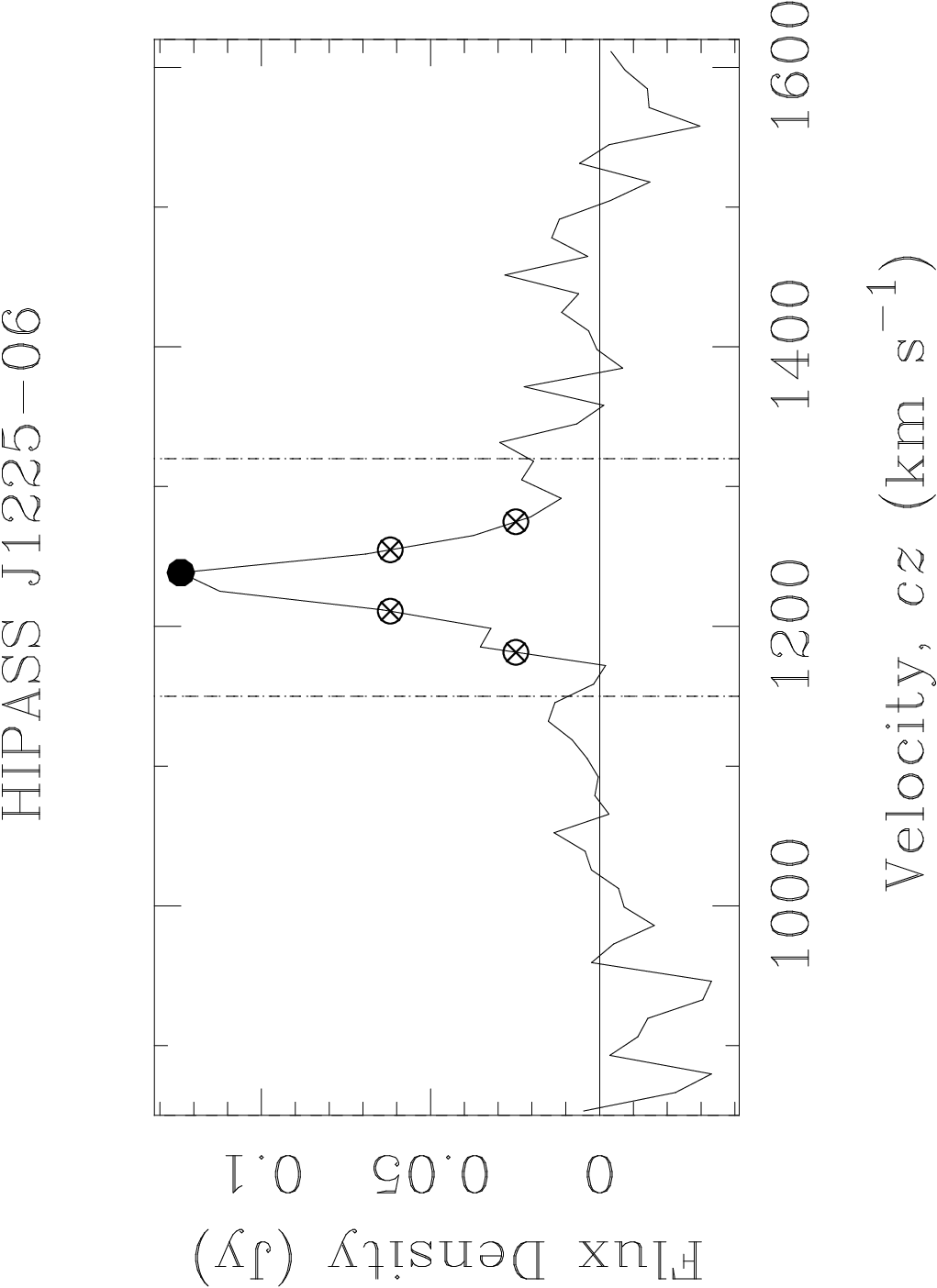}
\caption{\label{fig:J1225} a) \HI\ spectrum of HIPASS J1225--08. b)
  DSS image centered on HIPASS J1225--08. The contour levels are at
  60, 70, 80 and 90\% of the maximum \HI\ flux density (5.2
  Jy\,beam$^{-1}$ \kms). There are two potential optical counterparts
  to HIPASS J1244--08 (see Table~\ref{tab:gal}).}
\end{figure}

\begin{figure}  
\vspace{5cm}
\includegraphics{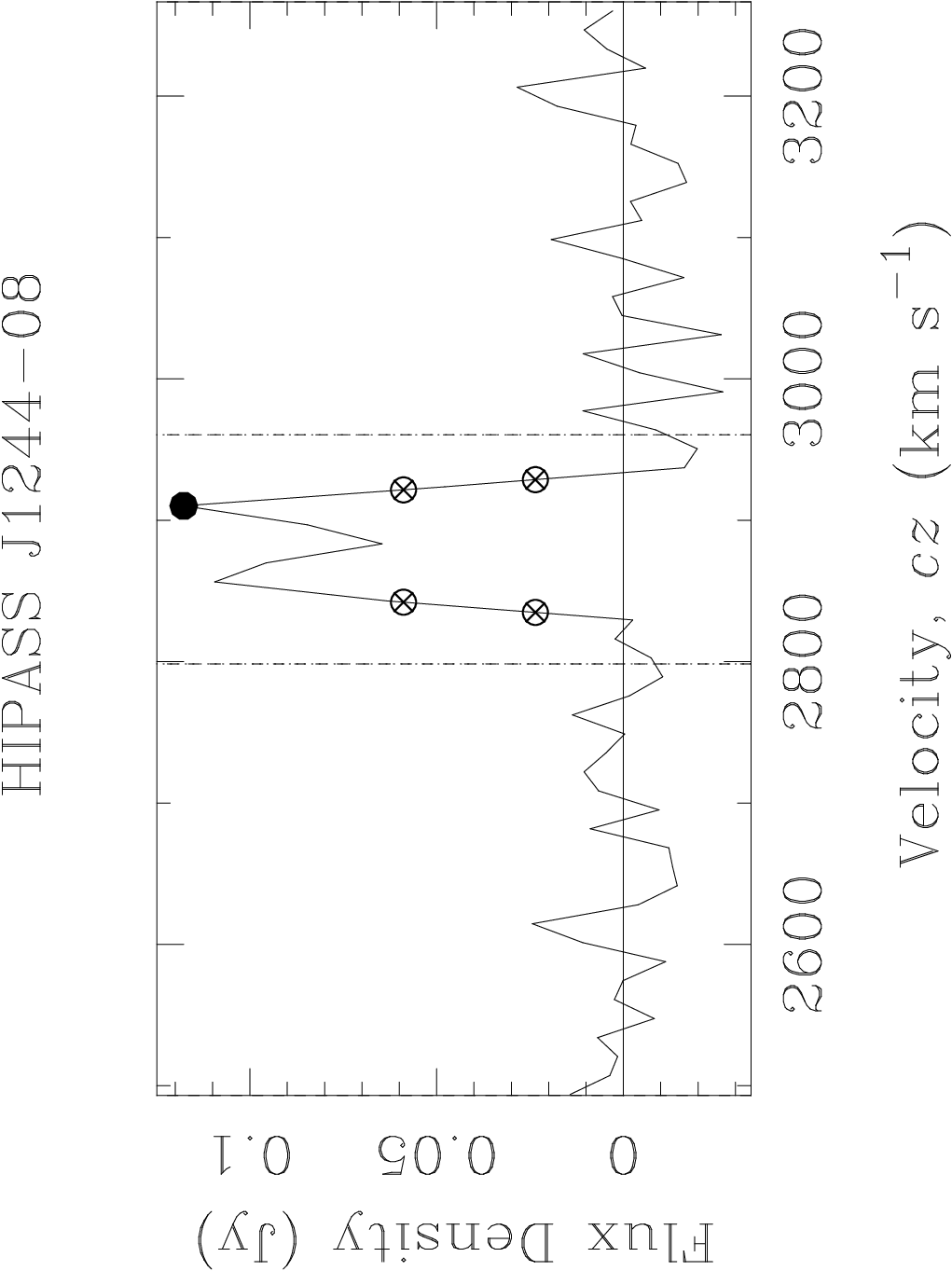}
\caption{\label{fig:J1244} a) \HI\ spectrum of HIPASS J1244--08. b)
  DSS image centered on HIPASS J1244--08. The contour levels are at
  60, 70, 80 and 90\% of the maximum \HI\ flux density (5.2
  Jy\,beam$^{-1}$ \kms). There are at least four potential optical
  counterparts to HIPASS J1244--08 (see Table~\ref{tab:gal}).}
\end{figure}

\begin{figure}  
\vspace{5cm}
\includegraphics{ryan-weber.fig9a.ps}
\caption{\label{fig:J1647} a) \HI\ spectrum of HIPASS J1647--00. b)
  DSS image centered on HIPASS J1647--00. The contour levels are at
  60, 70, 80 and 90\% of the maximum \HI\ flux density (10.0
  Jy\,beam$^{-1}$ \kms). There are three potential optical
  counterparts to HIPASS J1647--00 (see Table~\ref{tab:gal}).}
\end{figure}

\begin{figure}  
\vspace{8cm}
\includegraphics{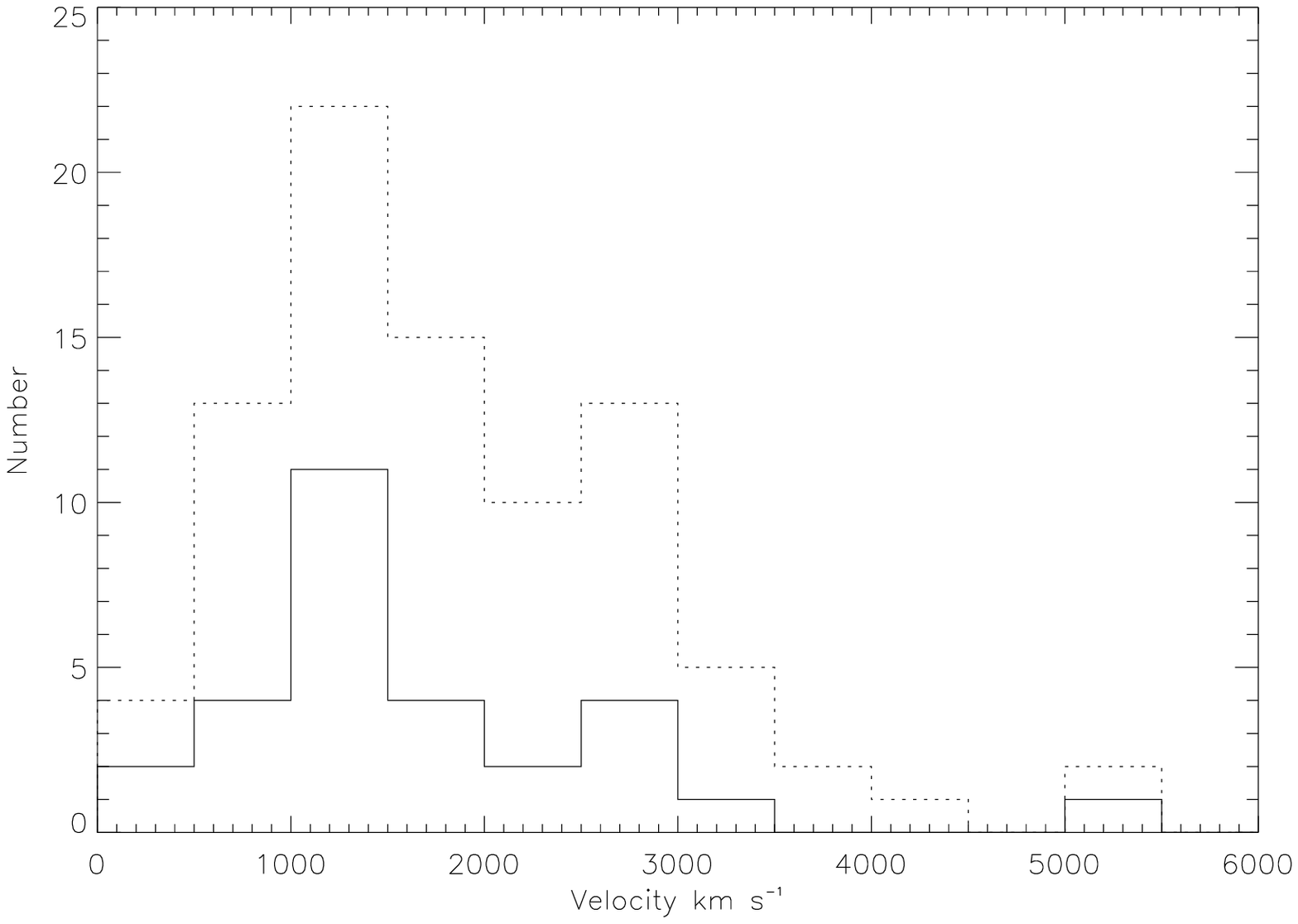}
\caption{\label{fig:vel}\HI\ velocity distribution of the newly
  cataloged galaxies in the HIPASS Bright Galaxy Catalog (dotted
  histogram). The solid histogram shows the newly cataloged galaxies
  with \bgt.}
\end{figure}

\begin{figure}  
\vspace{8cm}
\includegraphics{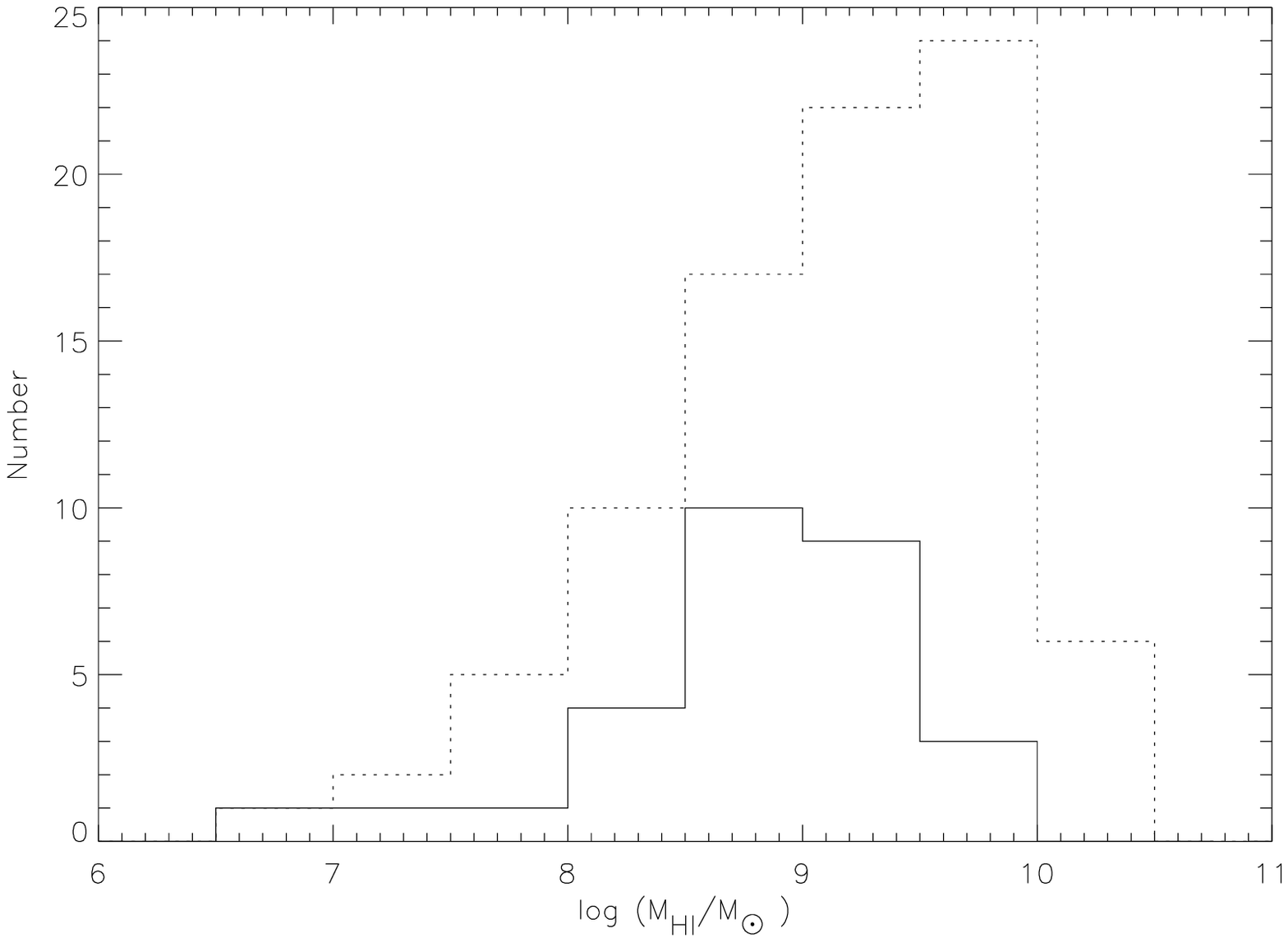}
\caption{\label{fig:mass}\HI\ mass distribution of the newly
  cataloged galaxies in the HIPASS Bright Galaxy Catalog (dotted
  histogram). The solid histogram shows the newly cataloged galaxies
  with \bgt.}
\end{figure}

\begin{figure}  
\vspace{8cm}
\includegraphics{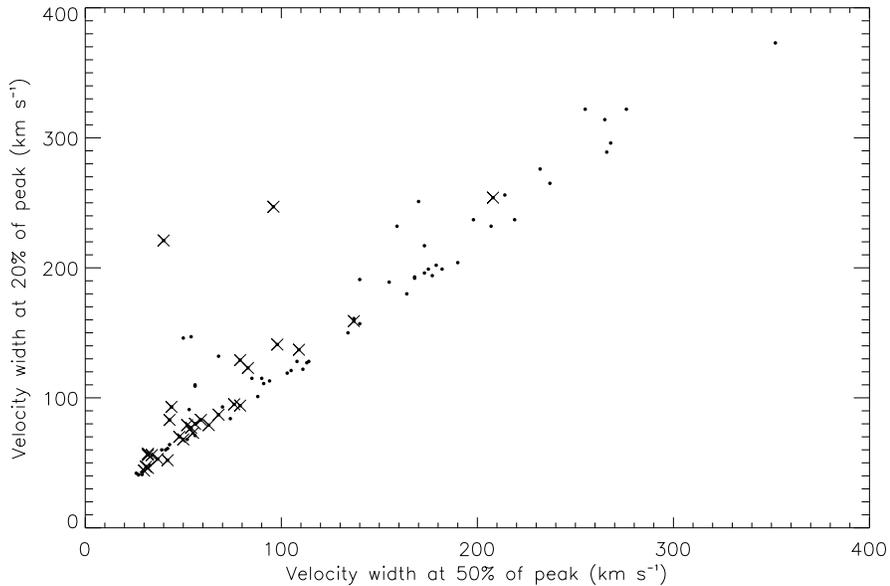}
\caption{\label{fig:curve} Comparison of the measured 50\% \HI\
  velocity width, \wfi, versus the 20\% \HI\ velocity width, \wtw, for
  all the newly cataloged galaxies from the HIPASS Bright Galaxy
  Catalog. Newly Cataloged galaxies with \bgt\ are also marked
  ($\times$).}
\end{figure}

\begin{figure}  
\caption{\label{fig:atca} ATCA \HI\ contours overlaid on 10\arcmin\
$\times$ 10\arcmin\ SuperCOSMOS fields for the galaxies HIPASS
J0705--20 ($b$ = -6\degr.4), J1430--54 ($b$ = 5\degr.5), J1434--47
($b$ = 12\degr.1), J1436--53 ($b$ = 6\degr.1), J1451--50 ($b$ =
8\degr.2), and J1506--49 ($b$ = 7\degr.7), arranged from top left to
bottom right.  These galaxies were observed on 2001 October 12, with
the ATCA EW352 compact configuration (integration time $\sim$100
minutes each).  The ATCA beam is displayed in the bottom left corner
of each image. The \HI\ contours levels are 0.5, 1, 2 then increasing
in increments of 1 Jy\,beam$^{-1}$ \kms.}
\end{figure}

\begin{figure}  
\caption{\label{fig:hispectra2} \HI\ spectra of the newly cataloged
  galaxies with low absolute Galactic latitudes (\blt) in the HIPASS
  BGC.}
\end{figure}

\setcounter{figure}{13}
\begin{figure}  
\caption{\label{fig:hispectra2cont} continued. }
\end{figure}

\clearpage
\pagestyle{empty}

\begin{table} 
\begin{tabular}{p{35mm}|p{45mm}|p{35mm}|p{20mm}}
\hline
Category           & Number $|b| < 10\degr$ & Number \bgt   & Total \\
\hline
\hline
\multicolumn{4}{l}{Newly Cataloged Galaxies (no optical or infrared 
counterpart in NED)} \\
\hline
Single counterpart &           13           &       25      &  38\\
Confused           &            1           &        4      &   5\\
No optical seen    &           43           & 1 (Behind LMC)&  44\\
\hline
Total              &           57 (including 33 HIZSS)& 30   &  87\\
\hline
\multicolumn{4}{l}{Galaxies with new redshifts (measured for the first 
time by the Parkes \HI\ multibeam surveys)}\\
\hline
                   &  17           & 34            &  51\\
\hline
\hline
Total              &               &               & 138\\
\hline
\end{tabular}
\caption{\label{tab:numbers} Number distribution of HIPASS BGC galaxies
  presented in this paper.}
\end{table}

\clearpage

\begin{landscape}
{\scriptsize     
\begin{longtable}{l|ccrrrrrrr|ccl|l} 
\caption{\HI\ properties of the 87 newly cataloged galaxies in the
HIPASS Bright Galaxy Catalog. Optical properties are given for those
\HI\ sources with one or more counterparts in the Digitized Sky Survey
(DSS).} \\
\hline
&\multicolumn{9}{c}{HIPASS Properties}&\multicolumn{3}{c}{Optical Properties}\\
HIPASS Name & \multicolumn{2}{c}{$\alpha,\delta$(J2000)}     & $l$   & $b$  
            & \vsys & \wfi & \wtw & \fhi & log \mhi
	    & \multicolumn{2}{c}{$\alpha,\delta$(J2000)} 
	    & Morphology & References \\
            & [$^{\rm h, m, s}]$ & [\degr\,\arcmin\,\arcsec] &[\degr]&[\degr] 
	    & [\tiny{\kms}] & [\tiny{\kms}] & [\tiny{\kms}] 
            & [\tiny{Jy\,\kms}] & [\msun]
	    & [$^{\rm h, m, s}]$ & [\degr\,\arcmin\,\arcsec] 
	    &            & \\
\hline
\hline
\endhead
\hline
continued\\
\endfoot
\\
\endlastfoot
\hline
HIPASS J0255--10 & 02:55:27 & --10:49:28 & 189.7 &  --56.6  & 1568 &  98 & 141  &   9.9 &   8.99  &02:55:19 &--10:49:14 &Im/BCD  &\\
HIPASS J0403--01 & 04:03:36 & --01:57:26 & 192.7 &  --37.6  &  910 &  96 & 247  &  16.1 &   8.70  &04:03:33 &--01:55:45 &Im      &\\
HIPASS J0447--57 & 04:47:14 & --57:12:35 & 266.2 &  --39.2  & 1248 &  43 &  83  &   7.6 &   8.52  &04:47:14 &--57:08:30 &Im      &\\
HIPASS J0532--67 & 05:31:44 & --67:21:33 & 277.5 &  --32.5  & 1375 &  55 &  73  &   6.9 &   8.56  &05:31:49 &--67:21:34 &Sa/Sb   &J0532--67 [5], [6]\\
HIPASS J0546--68 & 05:46:21 & --68:41:38 & 278.9 &  --31.0  & 1306 &  26 &  42  &   4.6 &   8.33  &         &           &        &J0546--68 [5] \\
HIPASS J0605--14 & 06:05:07 & --14:06:06 & 220.6 &  --16.5  & 3062 &  40 & 221  &  13.8 &   9.68  &06:04:59 &--14:06:29 &Im      &\\
&&&&&&&&& &06:05:11 &--14:02:10 &Sm      &\\		    			          
&&&&&&&&& &06:05:21 &--14:03:42 &BCD     &\\		    			          
HIPASS J0617--17 & 06:17:52 & --17:08:50 & 224.8 &  --15.0  &  855 &  32 &  56  &  10.3 &   8.26  &06:17:53 &--17:09:04 &Im/BCD  &\\
HIPASS J0700--04 & 07:00:29 & --04:11:37 & 217.7 &     0.1  &  298 &  70 &  93  &  26.6 &   7.17  &          &          &        &HIZSS003 [3] \\
HIPASS J0705--20 & 07:05:45 & --20:59:30 & 233.3 &   --6.4  &  766 &  29 &  41  &  13.2 &   8.19  &          &          &        &\\
HIPASS J0718--09 & 07:18:25 & --09:02:46 & 224.1 &     1.8  &  916 &  53 &  91  &  14.2 &   8.47  &07:18:21 &--09:03:20 &Sd/Sm   &HIZSS006 [3] \\
&&&&&&&&&        &07:18:15 &--09:03:00 &Sd/Sm   &HIZSS006 [3] \\			 	 
HIPASS J0730--22 & 07:30:08 & --22:01:27 & 236.8 &   --1.9  &  779 & 268 & 296  &  86.4 &   9.00  &07:30:08 &--22:01:06 &Scd/Sd  &HIZSS012 [3], [6]\\
HIPASS J0733--28 & 07:33:16 & --28:41:07 & 243.0 &   --4.4  & 2091 &  54 & 147  &  10.8 &   9.18  &         &           &        &HIZSS013 [3] \\
HIPASS J0736--19 & 07:36:09 & --19:25:18 & 235.2 &     0.6  &  786 &  56 &  71  &   6.5 &   7.90  &         &           &        &HIZSS014 [3] \\
HIPASS J0742--34 & 07:42:45 & --34:38:21 & 249.2 &   --5.5  & 2898 & 232 & 276  &  33.5 &   9.98  &07:42:38 &--34:38:28 &Sc/Sd   &HIZSS019 [3], [6]\\
HIPASS J0744--35 & 07:44:14 & --35:48:53 & 250.4 &   --5.9  & 2879 & 265 & 314  &  26.4 &   9.87  &07:44:12 &--35:48:34 &Sc 	 &\\
HIPASS J0746--28 & 07:46:21 & --28:27:51 & 244.2 &   --1.8  &  494 &  85 & 115  &  22.5 &   7.68  &07:46:16 &--28:28:10 &Im	&HIZSS021 [3] \\
HIPASS J0749--35 & 07:49:31 & --35:41:34 & 250.8 &   --4.9  & 2865 &  43 &  64  &  15.0 &   9.62  & & &     & HIZSS025 [3]  \\
HIPASS J0751--37 & 07:51:27 & --37:12:56 & 252.4 &   --5.3  & 2804 &  88 & 101  &   9.6 &   9.41  & & & 	     & 	      \\
HIPASS J0751--55 & 07:51:30 & --55:28:00 & 268.5 &  --14.2  & 1119 &  42 &  52  &   6.2 &   8.25  &07:51:23 &--55:27:13 & Sm/Im  &[KK2000] 24 [7] \\
HIPASS J0806--37 & 08:06:59 & --37:43:17 & 254.4 &   --2.9  &  860 &  56 & 109  &   9.8 &   8.13  & & &     & HIZSS033 [3]  \\
HIPASS J0826--44 & 08:26:26 & --44:19:28 & 261.9 &   --3.6  & 1023 & 182 & 199  &  37.1 &   8.91  & & &     & HIZSS043 [3] \\
HIPASS J0833--37 & 08:33:56 & --37:32:51 & 257.3 &     1.6  &  958 &  56 & 110  &   9.5 &   8.25  &08:34:00 &--37:32:59 & Sm?	&HIZSS045 [3] \\
HIPASS J0834--40 & 08:34:41 & --40:08:45 & 259.4 &     0.1  & 2771 & 168 & 192  &  25.6 &   9.82  &         &           &        &HIZSS046 [3] \\
HIPASS J0902--40 & 09:02:02 & --40:06:37 & 262.7 &     4.2  & 1636 &  94 & 113  &  12.0 &   8.96  &         &           &        &HIZSS051 [3] \\
HIPASS J0904--37 & 09:04:36 & --37:22:35 & 261.0 &     6.4  & 1033 & 105 & 121  &  12.0 &   8.44  &09:04:42 &--37:22:20 &Sc/Sd 	 &\\
HIPASS J0917--53 & 09:17:36 & --53:22:39 & 274.3 &   --2.9  &  946 & 173 & 217  &  18.8 &   8.52  &09:17:31 &--53:23:19 &Sc	 &HIZSS053 [3] \\
HIPASS J0927--55 & 09:27:47 & --55:59:09 & 277.1 &   --3.7  & 1156 & 140 & 157  &  34.7 &   9.03  &         &           &        &HIZSS054 [3] \\
HIPASS J0949--56 & 09:49:36 & --56:31:20 & 279.8 &   --2.1  & 1762 & 214 & 256  &  42.3 &   9.58  &         &           &        &HIZSS059 [3] \\
HIPASS J0957--48 & 09:57:03 & --48:55:41 & 275.9 &     4.6  & 3727 &  33 &  48  &   7.3 &   9.56  &09:57:11 &--48:56:30 & Spiral &HIZSS060 [3] \\
HIPASS J1004--73 & 10:04:07 & --73:50:50 & 291.9 &  --14.7  & 1246 &  50 &  68  &   8.0 &   8.51  &10:04:58& --73:51:19 &SBm	&J1005--73 [5]\\
HIPASS J1015--34 & 10:15:47 & --34:05:21 & 269.7 &    18.5  & 2608 &  32 &  57  &   5.7 &   9.11  &10:15:38 &--34:06:11 &BCD       &\\
HIPASS J1053--62 & 10:53:44 & --62:50:43 & 290.0 &   --3.0  & 1836 & 266 & 289  &  33.5 &   9.53  &         &           &        &HIZSS066 [3], [5] \\
HIPASS J1101--65 & 11:01:53 & --65:45:32 & 292.0 &   --5.2  & 1790 &  42 &  61  &   8.8 &   8.93  &         &           & 	&J1101--65 [5]\\
HIPASS J1106--14 & 11:06:05 & --14:22:10 & 268.1 &    41.3  & 1040 &  76 &  95  &  11.2 &   8.49  &11:06:11 &--14:24:28 &Im      &[KKS2000] 23 [8] \\
HIPASS J1118--17 & 11:18:03 & --17:38:26 & 273.5 &    39.8  & 1069 &  52 &  79  &  10.2 &   8.48  &11:18:03 &--17:38:32 &BCD     &\\
HIPASS J1141--64 & 11:41:19 & --64:28:41 & 295.5 &   --2.6  & 2025 & 179 & 202  &  38.6 &   9.70  &         &           &        &HIZSS068 [3], [5] \\
HIPASS J1149--64 & 11:49:50 & --64:00:24 & 296.2 &   --1.9  & 2067 & 255 & 322  &  45.7 &   9.79  &         &           &        &HIZSS069 [3], [5] \\
HIPASS J1202--61 & 12:02:56 & --61:39:03 & 297.2 &     0.7  & 1540 & 207 & 232  &  93.7 &   9.80  &         &           &        &HIZSS070 [3] \\
HIPASS J1204--63 & 12:04:20 & --63:11:27 & 297.6 &   --0.8  & 2034 & 168 & 193  &  29.3 &   9.58  &         &           &        &HIZSS071 [3], [5] \\
HIPASS J1221--59 & 12:21:39 & --59:42:21 & 299.2 &     2.9  & 1477 & 175 & 199  &  40.1 &   9.40  &         &           &        &HIZSS073 [3] \\
HIPASS J1225--06 & 12:25:33 & --06:31:09 & 291.4 &    55.8  & 1233 &  44 &  93  &   7.3 &   8.55  &12:25:51 &--06:29:22 &Im/BCD  &[9]\\
&&&&&&&&&        &12:25:39 &--06:33:08 &Im/BCD  &\\				 	 
HIPASS J1244--08 & 12:44:59 & --08:18:23 & 300.2 &    54.5  & 2882 &  79 &  94  &   7.3 &   9.36  &12:45:13 &--08:21:31 &Sm      &\\
&&&&&&&&&        &12:45:08 &--08:23:05 &Sm      &[6]\\				 	 
&&&&&&&&&        &12:45:04 &--08:23:46 &Im/BCD  &[10]\\				 	 
HIPASS J1247--77 & 12:47:26 & --77:34:17 & 302.7 &  --14.7  &  413 &  32 &  46  &   4.7 &   6.75  &12:47:34 &--77:34:54 &Im      &[5]	\\
HIPASS J1248--08 & 12:48:28 & --08:01:49 & 301.7 &    54.8  & 1502 &  63 &  79  &  22.2 &   9.23  &12:48:31 &--08:02:37 &Sc	 &\\
HIPASS J1255--03 & 12:55:16 & --03:23:39 & 304.8 &    59.5  & 1484 &  34 &  56  &   8.0 &   8.79  &12:55:11 &--03:24:12 &Im	 &\\
HIPASS J1258--33 & 12:58:43 & --33:45:21 & 304.7 &    29.1  & 2476 &  59 &  83  &   7.5 &   9.21  &12:58:36 &--33:45:35 &SBm	 &\\
HIPASS J1300--13B& 13:00:58 & --13:31:27 & 306.5 &    49.3  & 1309 &  56 &  80  &   7.0 &  8.59	  &13:01:07 &--13:31:04 &SBm(pec)&\\
HIPASS J1312--60 & 13:12:47 & --60:52:40 & 305.5 &     1.9  & 2321 & 237 & 265  &  32.3 &   9.77  & & &     & HIZSS076 [3]  \\
HIPASS J1321--31 & 13:21:07 & --31:33:03 & 310.3 &    30.9  &  571 &  31 & 47  &    5.9 &  7.54  & 13:21:08&  --31:31:45&  Im    &[2], [KK98] 195 [11] \\
HIPASS J1333--58 & 13:33:00 & --58:03:50 & 308.4 &    4.4  & 1476 & 111 & 122  &   12.4 &  8.90  & & &                           &HIZSS080 [3], [4] \\
HIPASS J1337--39 & 13:37:30 & --39:52:56 & 312.5 &   22.1  &  492 &  37 &  53  &    6.6 &  7.36  &13:37:26 &--39:53:47 & Im      &[2] \\
HIPASS J1415--04A& 14:15:08 & --04:20:00 & 338.8 &   52.6  & 2740 & 208 & 254  &   21.4 &  9.81  &14:15:17 &--04:21:31 &SBd      &[6], [12] \\
HIPASS J1415--04B& 14:15:56 & --04:04:02 & 339.3 &   52.7  & 2730 &  54 &  77  &    8.0 &  9.38  &14:15:47 &--04:04:32 &SBb/c    &[12] \\
HIPASS J1424--16B& 14:24:29 & --16:58:58 & 332.7 &   40.5  & 1487 &  68 &  87  &   13.4 &  9.03  &14:24:31&--16:59:15 &Sm/Im     &\\
HIPASS J1430--54 & 14:30:18 & --54:36:23 & 317.0 &    5.5  & 3020 & 114 & 128  &    9.6 &  9.50  &14:30:16 & --54:36:22&Sc 	     & 	      \\
HIPASS J1434--47 & 14:34:36 & --47:12:05 & 320.5 &   12.1  & 1512 &  30 &  44  &    4.8 &  8.55  &14:34:44& --47:13:35 &Im 	     & 	      \\
HIPASS J1436--53 & 14:36:50 & --53:34:40 & 318.3 &    6.1  & 3016 & 108 & 128  &   31.2 & 10.02  &14:36:48& --53:34:22  & Im     &WKK3285 [13]\\                
HIPASS J1441--62 & 14:41:37 & --62:44:38 & 315.2 &   --2.5  &  672 &  52 &  68  &   4.7 &   7.62 &        &             &        &[5]\\
HIPASS J1451--50 & 14:51:21 & --50:13:55 & 321.8 &    8.2   & 1275 & 164 & 180  &  24.8 &   9.09 &14:51:13 &-50:12:47 & Sm    & 	      \\
HIPASS J1501--60 & 15:01:30 & --60:44:53 & 318.2 &   --1.8  & 4436 & 134 & 150  &  14.7 &  10.04  & & &                          &HIZSS092 [3], [4]\\
HIPASS J1506--49 & 15:06:58 & --49:24:47 & 324.4 &     7.7  & 1041 & 113 & 127  &  29.7 &   8.97  & & &     & 	      \\
HIPASS J1513--44 & 15:13:10 & --44:03:10 & 328.1 &    11.8  & 5125 &  48 &  70  &   7.8 &   9.91  &15:13:13  &--44:02:00 &BCD/Im 	     & 	      \\
HIPASS J1522--49 & 15:22:22 & --49:22:09 & 326.6 &     6.4  & 2307 & 173 & 196  &  19.3 &   9.57  &15:22:24 &--49:21:29 &Im 	&WKK4860  [13]\\
HIPASS J1526--51 & 15:26:18 & --51:09:46 & 326.1 &     4.6  &  605 &  39 &  60  &   6.0 &   7.68  &         &           &        &HIZOA J1526--51 [4] \\
HIPASS J1532--56 & 15:32:55 & --56:08:35 & 324.1 &   --0.1  & 1363 &  68 & 132  &  64.2 &   9.58  &         &           &        &[1], HIZSS097 [3], [4] \\
HIPASS J1558--10 & 15:58:27 & --10:30:45 & 359.7 &    31.1  &  933 &  79 & 129  &  11.4 &   8.62  &15:58:20 &--10:32:16 &Sm/BCD	 &\\
HIPASS J1605--57 & 16:05:19 & --57:52:04 & 326.5 &   --4.1  & 2991 &  91 & 111  &  17.5 &   9.77  &16:05:22 &--57:51:43 &Spiral  &HIZSS101 [3], [4], [13] \\
HIPASS J1621--58 & 16:21:50 & --58:00:06 & 328.0 &   --5.8  & 1404 &  74 &  84  &   9.4 &   8.79  &         &           &        &\\
HIPASS J1624--42 & 16:24:54 & --42:29:35 & 339.4 &     4.8  & 2232 & 159 & 232  &  21.5 &   9.61  &         &           &        &HIZSS104 [3]  \\
HIPASS J1629--57 & 16:29:58 & --57:39:09 & 329.0 &   --6.3  & 2685 & 198 & 237  &  14.4 &   9.59  &         &           &        &\\
HIPASS J1639--56 & 16:39:41 & --56:52:35 & 330.4 &   --6.7  & 1468 & 103 & 119  &  18.9 &   9.14  &         &           &        &\\
HIPASS J1647--00 & 16:47:55 & --00:23:08 &  18.1 &    27.4  & 2347 &  83 & 123  &  11.3 &   9.45  &16:47:59 &--00:22:59 &Sm(Gpair)&\\
&&&&&&&&& &16:48:10 &--00:21:48 &Spiral  &\\					 	 
&&&&&&&&& &16:47:59 &--00:19:47 &Sd      &\\					 	 
HIPASS J1705--29 & 17:05:26 & --29:40:38 & 354.5 &     6.9  & 2677 & 177 & 194  &  19.1 &   9.75  &         &           &        &\\
HIPASS J1711--47 & 17:11:35 & --47:35:59 & 340.8 &   --4.8  & 2187 & 219 & 237  &  21.1 &   9.59  &         &           &        &HIZSS106 [3] \\
HIPASS J1719--41 & 17:19:48 & --41:18:14 & 346.8 &   --2.3  & 3902 & 276 & 322  &  27.0 &  10.22  &         &           & 	&HIZSS107 [3] \\
HIPASS J1758--31 & 17:58:37 & --31:18:11 & 359.4 &   --3.6  & 3316 &  41 &  60  &   5.5 &   9.40  & & &     & 	      \\
HIPASS J1807--02 & 18:07:10 & --02:49:17 &  25.3 &     8.5  & 1765 & 137 & 161  &  35.3 &   9.72  & & & 	   & 	      \\
HIPASS J1807--08 & 18:07:39 & --08:36:38 &  20.2 &     5.7  & 3485 & 170 & 251  &  19.0 &  10.01  & & &     & 	      \\
HIPASS J1812--21 & 18:12:22 & --21:34:07 &   9.4 &   --1.6  & 1533 &  50 & 146  &  11.2 &   9.07  & & &     & 	      \\
HIPASS J1824--01 & 18:24:58 & --01:28:03 &  28.6 &     5.2  & 2865 & 352 & 373  &  31.6 &  10.08  & & &     & 	      \\
HIPASS J1838--22 & 18:38:36 & --22:48:25 &  11.1 &   --7.5  & 1656 &  27 &  41  &   9.4 &   9.06  & & &     & 	      \\
HIPASS J1841--18 & 18:41:05 & --18:59:54 &  14.8 &   --6.3  & 1671 & 155 & 189  &  15.7 &   9.30  & & &     & 	      \\
HIPASS J1851--09 & 18:51:19 & --09:10:53 &  24.7 &   --4.1  & 5485 &  90 & 115  &   9.8 &  10.11  & & &     & 	      \\
HIPASS J1856--03 & 18:56:00 & --03:10:21 &  30.6 &   --2.5  & 1582 & 190 & 204  &  21.7 &   9.44  &         &           &        &HIZSS108 [3] \\
HIPASS J1901--04 & 19:01:44 & --04:29:37 &  30.1 &   --4.3  & 1530 & 140 & 191  &  23.9 &   9.45  &         &           &        &HIZSS109 [3] \\
HIPASS J2020--04 & 20:20:35 & --04:54:16 &  38.9 &  --22.0  & 1387 & 109 & 137  &  11.9 &   9.09  &20:20:32 &--04:54:00 &Sm/Im   &\\
HIPASS J2200--56 & 22:00:42 & --56:28:10 & 336.9 &  --47.9  & 1847 & 137 & 159  &  19.0 &   9.40  &22:00:40 &--56:28:20 &BCD     &[9]\\

\hline
\label{tab:gal}
\end{longtable}
\noindent Catalog and reference codes: 
  [1] Staveley-Smith et al. (1998); 
  [2] Banks et al. (1999);  
  [3] HIZSS, Henning et al. (2000);  
  [4] HIZOA, Juraszek et al. (2000); 
  [5] HIPASS SCC, Kilborn et al.  (2002); 
  [6] 2MASXi, Jarret et al. (2000); 
  [7] KK2000, Karachentseva \& Karachentsev (2000); 
  [8] KKS2000, Karachentsev et al.  (2000); 
  [9] APMUKS(BJ), Maddox et al. (1990); 
  [10] NPM1G, Klemola et al. (1987); 
  [11] KK98, Karachentseva \& Karachentsev (1998);
  [12] 2dFGRS, Colless et al. (2001); 
  [13] WKK, Woudt \& Kraan-Korteweg (2001). 
}
\end{landscape}

\clearpage
{\scriptsize
\begin{longtable}{lccrrcccccl} 
\caption{\HI\ properties of the additional 51 galaxies without
         a previous redshift measurement
         in the HIPASS Bright Galaxy Catalog. } \\
\hline
HIPASS Name & \multicolumn{2}{c}{$\alpha,\delta$(J2000)} 
            &   $l$   &   $b$   
	    & \vsys  & \wfi   & \wtw  & \fhi & log \mhi & NED-ID\\
            & [$^{\rm h, m, s}]$ & [ \degr\,\arcmin\,\arcsec ] 
            & [\degr] & [\degr]  
	    & [\tiny{\kms}] & [\tiny{\kms}] & [\tiny{\kms}]
            & [\tiny{Jy\,\kms}] & [\msun] &\\
\hline
\hline
\endhead
\hline
continued\\
\endfoot
\\
\endlastfoot
\hline
HIPASS J0136--60 & 01:36:22 & --60:23:41 & 293.1 &--55.9 & 2221 &  45 &  64 &   8.9 &  9.20 & ESO113-IG054   \\
HIPASS J0223--04 & 02:23:51 & --04:36:52 & 171.4 &--58.5 & 2273 &  96 & 112 &  13.9 &  9.49 & PGC009103      \\
HIPASS J0310--39 & 03:10:01 & --39:59:14 & 246.1 & -58.7 &  711 &  30 &  45 &   4.4 &  7.78 & ESO300-G016    \\
HIPASS J0348--39 & 03:48:32 & --39:26:46 & 243.1 &--51.4 & 1168 &  29 &  46 &   4.9 &  8.31 & ESO302-G?010   \\
HIPASS J0430--20 & 04:30:53 & --20:36:45 & 218.0 &--39.8 & 1627 & 122 & 149 &  18.7 &  9.24 & APMBGC551+05   \\
HIPASS J0439--47 & 04:39:51 & --47:30:14 & 253.7 &--41.5 & 1368 &  53 &  80 &   6.8 &  8.58 & ESO202-IG048   \\
HIPASS J0553--59 & 05:53:12 & --59:03:03 & 267.7 &--30.4 & 1308 & 120 & 141 &  14.4 &  8.82 & ESO120-G021    \\
HIPASS J0555--29 & 05:55:07 & --29:56:23 & 235.4 &--24.5 & 3604 &  31 &  47 &   5.9 &  9.45 & ESO424-G039    \\
HIPASS J0600--31 & 06:00:18 & --31:48:50 & 237.8 & -24.1 & 1353 &  90 & 117 &  10.0 &  8.72 & ESO425-G001    \\
HIPASS J0615--57 & 06:15:58 & --57:44:29 & 266.5 &--27.3 &  577 &  65 &  96 &  14.1 &  7.76 & ESO121-G020    \\
HIPASS J0649--14 & 06:49:38 & --14:25:08 & 225.6 & --6.9 & 2802 & 126 & 147 &  14.5 &  9.61 & CGMW1-0409     \\
HIPASS J0651--44 & 06:51:18 & --44:05:17 & 253.7 &--18.5 & 4748 &  46 & 153 &   8.4 &  9.85 & ESO256-G003    \\
HIPASS J0659--01 & 06:59:22 & --01:31:43 & 215.2 &   1.1 & 1733 & 168 & 188 &  20.2 &  9.31 & CGMW1-0476     \\
HIPASS J0712--28 & 07:12:42 & --28:40:21 & 240.9 & --8.4 &  881 & 112 & 128 &  10.9 &  8.25 & ESO428-G004    \\
HIPASS J0718--57 & 07:18:21 & --57:26:59 & 268.5 &--19.2 & 1148 &  64 &  84 &  10.9 &  8.53 & AM0717-571     \\
HIPASS J0725--17 & 07:25:55 & --17:53:15 & 232.7 & --0.8 & 2758 &  94 & 167 &  11.8 &  9.50 & CGMW1-0877c    \\
HIPASS J0726--09 & 07:26:34 & --09:14:24 & 225.2 &   3.5 & 2438 &  30 &  55 &   6.2 &  9.11 & ZOAG\_G225+03  \\
HIPASS J0735--50 & 07:35:18 & --50:15:57 & 262.6 &--14.0 & 1200 &  91 & 123 &  13.0 &  8.66 & ESO208-G025    \\
HIPASS J0741--38 & 07:41:30 & --38:35:39 & 252.6 & --7.7 & 2789 &  25 &  44 &  13.1 &  9.54 & ESO311-G003    \\
HIPASS J0747--26 & 07:47:02 & --26:21:13 & 242.5 & --0.6 &  881 & 141 & 153 &  45.7 &  8.86 & IRAS07451--2610\\
HIPASS J0807--17 & 08:07:00 & --17:28:46 & 237.3 &   7.9 & 2370 & 138 & 164 &  20.3 &  9.58 & CGMW2-2253     \\
HIPASS J0809--41 & 08:09:54 & --41:41:03 & 258.0 & --4.6 & 1995 & 312 & 337 &  25.7 &  9.49 & IRAS08081--4132\\
HIPASS J0857--29 & 08:57:09 & --29:09:29 & 253.6 &  10.5 & 1971 &  20 &  37 &   3.3 &  8.60 & CGMW2-4513     \\
HIPASS J0857--39 & 08:57:28 & --39:16:04 & 261.5 &   4.1 &  978 & 317 & 349 &  37.0 &  8.86 & ESO314-G?002   \\
HIPASS J0926--60 & 09:26:26 & --60:35:39 & 280.2 &  -7.1 & 2120 & 125 & 160 &  18.6 &  9.42 & ESO126-G011c   \\
HIPASS J0945--33 & 09:45:33 & --33:48:07 & 264.4 &  14.8 & 2654 &  63 & 117 &   7.9 &  9.27 & ESO373-IG022   \\
HIPASS J0953--61 & 09:53:00 & --61:30:14 & 283.3 &  -5.7 & 4439 &  48 &  68 &   7.3 &  9.72 & RKK1733        \\
HIPASS J0957--39 & 09:57:15 & --39:00:10 & 269.7 &  12.4 & 4652 &  54 &  99 &   8.2 &  9.82 & ESO316-G006    \\
HIPASS J1003--26B& 10:03:51 & --26:38:33 & 262.6 &  22.8 &  885 &  77 & 101 &   9.5 &  8.17 & ESO499-G038c   \\
HIPASS J1005--28 & 10:05:33 & --28:23:40 & 264.1 &  21.7 & 1037 & 100 & 195 &  18.8 &  8.66 & ESO435-G039c   \\
HIPASS J1013--34 & 10:13:07 & --34:54:50 & 269.7 &  17.5 & 4367 &  43 & 102 &   9.5 &  9.82 & ESO374-G043    \\
HIPASS J1040--54 & 10:40:20 & --54:32:31 & 284.6 &   3.6 & 2753 & 109 & 138 &  15.3 &  9.59 & RKK2791/89     \\
HIPASS J1041--48 & 10:41:25 & --48:19:40 & 281.7 &   9.1 &  989 &  70 &  80 &  12.3 &  8.40 & ESO214-G018c   \\
HIPASS J1057--48 & 10:57:32 & --48:11:02 & 284.1 &  10.5 &  598 &  67 &  83 & 104.4 &  8.63 & ESO215-G?009   \\
HIPASS J1126--72 & 11:26:15 & --72:37:06 & 296.6 &--10.8 & 2031 &  28 &  38 &  12.2 &  9.20 & PGC035171      \\
HIPASS J1227--34 & 12:27:45 & --34:25:08 & 297.4 &  28.2 & 2922 & 120 & 196 &  12.9 &  9.59 & ESO380-IG033   \\
HIPASS J1305--28 & 13:05:52 & --28:22:11 & 306.8 &  34.4 & 2282 & 105 & 170 &  17.8 &  9.51 & ESO443-G061    \\
HIPASS J1329--48 & 13:29:05 & --48:09:57 & 309.4 &  14.2 & 2034 & 125 & 151 &  12.3 &  9.23 & ESO220-G014    \\
HIPASS J1338--56 & 13:38:10 & --56:28:30 & 309.4 &   5.8 & 3957 & 312 & 342 &  25.6 & 10.17 & PGC048178      \\
HIPASS J1343--44 & 13:43:07 & --44:51:10 & 312.5 &  17.1 & 2200 & 107 & 154 &  12.0 &  9.30 & ESO270-G026    \\
HIPASS J1403--27 & 14:03:31 & --27:17:09 & 322.0 &  32.9 & 1327 & 117 & 131 &  12.0 &  8.84 & ESO510-IG052   \\
HIPASS J1409--51 & 14:09:05 & --51:10:43 & 315.1 &   9.8 & 4530 & 132 & 164 &  14.1 & 10.04 & ESO221-G028    \\
HIPASS J1517--43 & 15:17:46 & --43:29:01 & 329.1 &  11.8 & 5001 &  28 &  67 &   4.6 &  9.66 & ESO274-G009    \\
HIPASS J1539--41 & 15:39:39 & --41:10:54 & 333.9 &  11.4 & 2390 & 161 & 181 &  20.8 &  9.65 & ESO329-G?013   \\
HIPASS J1609--60 & 16:09:44 & --60:18:00 & 325.2 &  -6.3 & 3246 & 170 & 242 &  21.7 &  9.94 & ESO136-G020    \\
HIPASS J1722--05 & 17:22:22 & --05:43:07 &  17.1 &  16.9 & 1625 & 134 & 191 &  30.1 &  9.57 & IRAS17197--0538\\
HIPASS J1937--52 & 19:37:37 & --52:00:42 & 346.0 &--28.0 & 3157 & 206 & 247 &  23.6 &  9.98 & IC4877/5       \\
HIPASS J2100--71 & 21:00:13 & --71:48:32 & 321.9 &--35.5 & 2821 &  60 &  92 &   7.6 &  9.36 & IC5069         \\
HIPASS J2118--09 & 21:18:31 & --09:01:16 &  42.3 &--36.7 & 2574 &  59 &  79 &  16.9 &  9.72 & IRAS21158--0914\\
HIPASS J2145--49 & 21:45:16 & --49:02:00 & 348.5 &--48.2 & 1600 &  71 & 115 &  27.1 &  9.44 & ESO236-G039c   \\
HIPASS J2217--45 & 22:17:23 & --45:33:11 & 351.5 &--54.4 & 3792 &  36 &  58 &   5.0 &  9.47 & ESO289-G012    \\
\hline
\label{tab:gal2}
\end{longtable}
}

\clearpage

\begin{table} 
\begin{tabular}{l|ccccr}
\hline
HIPASS Name     &\multicolumn{2}{c}{ATCA $\alpha,\delta$(J2000)} &Offset from    &PA        &ATCA-\fhi\\
                &[$^{\rm h, m, s}]$ & [\degr\,\arcmin\,\arcsec]   &HIPASS [\arcmin]
                &[\degr]   &[Jy \kms]\\
\hline
\hline
HIPASS J0705--20   & 07:05:47&--20:59:30  &0.5     &100        &13.5\\
HIPASS J1430--54   & 14:30:17&--54:36:10  &0.3     &315--345    &7.3\\
HIPASS J1434--47   & 14:34:43&--47:13:30  &1.9     &---       &2.7\\  
HIPASS J1436--53   & 14:36:49&--53:34:27  &0.3     &100        &26.4\\
HIPASS J1451--50   & 14:51:13&--50:12:47  &1.7     &300        &23.7\\
HIPASS J1506--49   & 15:06:59&--49:25:39  &0.9     &160--170    &25.9\\
\hline
\end{tabular}
\caption{\label{table:atca} \HI\ parameters from ATCA observations of
six galaxies shown in Figure~\ref{fig:atca}. The position angles 
(PA) were derived from the velocity field and may be affected by the 
elongated beam.}
\label{tab:atca}
\end{table}

\end{document}